\documentclass[structabstract]{aa}  
\usepackage{graphicx}
\usepackage{natbib}
\usepackage{lscape}
\usepackage{threeparttable}
\usepackage{amsmath}
\usepackage{amssymb}
\usepackage{array}
\usepackage{txfonts}
\usepackage{arydshln}
\usepackage{hyperref}

\usepackage{color}

\authorrunning{Guti\'errez et al.}
\titlerunning{SN~2010ev: A reddened HVG SN.}

\begin{document}
\title{Supernova 2010ev: A reddened high velocity gradient type Ia supernova\thanks{This paper includes 
data gathered with the Du Pont Telescope at Las Campanas Observatory, Chile; and the
Gemini Observatory, Cerro Pachon, Chile (Gemini Program GS-2010A-Q-14). Based on observations collected at the European 
Organisation for Astronomical Research in the Southern Hemisphere, Chile (ESO Programme 085.D-0577)}}

\author{Claudia P. Guti\'errez\inst{1,2,3}
\and Santiago Gonz\'alez-Gait\'an \inst{1,2}
\and Gast\'on Folatelli\inst{4}
\and Giuliano Pignata\inst{5,1}
\and Joseph P. Anderson\inst{3} 
\and Mario Hamuy\inst{2,1}
\and Nidia Morrell\inst{6}
\and Maximilian Stritzinger\inst{7}
\and Stefan Taubenberger\inst{8,9}
\and Filomena Bufano\inst{1,5,10}
\and Felipe Olivares E.\inst{1,5}
\and Joshua B. Haislip\inst{11}
\and Daniel E. Reichart\inst{11}
}

\institute{Millennium Institute of Astrophysics, Casilla 36-D, Santiago, Chile, 
\and
Departamento de Astronom\'ia, Universidad de Chile, Casilla 36-D, 
Santiago, Chile
\and
European Southern Observatory, Alonso de C\'ordova 3107, Casilla 19, Santiago, Chile\\
\email{cgutierr@das.uchile.cl}
\and 
Instituto de Astrof\'isica de La Plata (IALP, CONICET), Argentina
\and 
Departamento de Ciencias Fisicas, Universidad Andres Bello, Avda. Rep\'ublica 252, Santiago, Chile
\and
Carnegie Observatories, Las Campanas Observatory, Casilla 601, La Serena, Chile
\and
Department of Physics and Astronomy, Aarhus University, Ny Munkegade 120, DK-8000 Aarhus C, Denmark
\and 
Max-Planck-Institut f\"ur Astrophysik, Karl-Schwarzschild-Str. 1, 85741 Garching, Germany
\and
European Southern Observatory, Karl-Schwarzschild-Str. 2, 85748 Garching, Germany
\and 
INAF - Osservatorio Astrofisico di Catania, Via Santa Sofia, 78, 95123, Catania, Italy
\and
University of North Carolina at Chapel Hill, Campus Box 3255, Chapel Hill, NC 27599-3255, USA
}

 
\abstract
{}
{We present and study the spectroscopic and photometric evolution of the type Ia supernova (SN~Ia) 2010ev.}
{We obtain and analyze multi-band optical light curves and optical/near-infrared spectroscopy at low and medium resolution 
spanning from $-$7 days to $+$300 days from the $B$-band maximum.}
{A photometric analysis shows that SN~2010ev is a SN Ia of normal brightness with a light curve shape of 
$\Delta m_{15}(B)=1.12 \pm 0.02$ and a stretch $s=0.94\pm 0.01$ suffering significant reddening. 
From photometric and spectroscopic analysis, we deduce a color excess of $E(B-V)=0.25\pm0.05$ and a 
reddening law of $R_v=1.54\pm0.65$. Spectroscopically, SN~2010ev belongs to the broad-line SN~Ia 
group, showing stronger than average $\ion{Si}{ii}$ $\lambda6355$ absorption features.  We also find that SN~2010ev is 
a high-velocity gradient SN, with $\dot v_{\mathrm{Si}}=164\pm7$ km s$^{-1}$ d$^{-1}$. The photometric
and spectral comparison with other supernovae shows that SN~2010ev has similar colors and velocities 
to SN~2002bo and SN~2002dj. The analysis of the nebular spectra indicates that the $\ion{[Fe}{ii]}$ $\lambda7155$
and $\ion{[Ni}{ii]}$ $\lambda7378$ lines are redshifted, as expected for a high velocity gradient supernova.
All these common intrinsic and extrinsic properties of the high velocity gradient (HVG) group
are different from the low velocity gradient (LVG) normal SN~Ia
population and suggest significant variety in SN~Ia explosions.}
{}


\maketitle
%

\section{Introduction}

Type Ia supernovae (SNe Ia) play an important role in stellar evolution and in the chemical 
enrichment of the universe, as well as in the determination of extragalactic distances, 
thanks to the relation between the decline rate of the light curve and its peak luminosity
\citep{Phillips93, Hamuy96, Phillips99} and between color and peak luminosity \citep{Tripp98}.
SNe Ia represent a homogeneous class and are thought to arise from the thermonuclear explosion 
of a carbon-oxygen white-dwarf either triggered by the interaction with the companion 
in a close binary system \citep{Hoyle60} or by direct collisions of white dwarfs. \citep{Raskin09}. 
In the leading scenario of a close binary system, the nature of the explosion 
and of the companion star  are still debated.
 Two of the models considered are: the single 
degenerate (SD) \citep{Nomoto82, Iben84},
and the double degenerate (DD) scenario \citep{Iben84, Webbink84}. In the former, a white dwarf 
accretes matter from the companion which can be a sub-giant or main sequence star, while
in the latter the SN is produced by the merging of two white dwarfs. SNe~Ia are thought to 
explode near the Chandrasekhar mass, although recent simulations of
sub-Chandrasekhar mass explosions have been successful for both scenarios \citep{Sim12,Kromer10,Pakmor12}.\\
\indent The study of SN~Ia spectral and photometric parameters in both early and late epochs can 
give key indications about the nature of the explosion. 
Studies of SN~Ia spectroscopic properties reveal significant diversity among the population.
\citet{Benetti05} defined a sub-classification of SNe~Ia based on
expansion velocities, line ratios and light curve decline rates. They classified the SN~Ia population
in three different sub-groups: High Velocity Gradient (HVG), Low Velocity Gradient (LVG) and FAINT
objects. A parallel classification was proposed by \citet{Branch06} based on absorption equivalent 
widths of $\ion{Si}{ii}$ $\lambda5972$ and $\lambda6355$ lines at maximum, which defines four
subtypes: Core-Normal (CN), Broad-Line (BL), Cool (CL) and Shallow Silicon (SS).  
\citet{Wang09} classified their SNe~Ia sample in two groups based on the blueshifted
velocity of $\ion{Si}{ii}$ absorption lines at maximum:
Normal velocity (NV; $v\sim10500$ km s$^{-1}$) and High velocity (HV; $v\geq12000$ km s$^{-1}$) SNe. 
Contemporary analyses of large samples of SNe~Ia spectra \citep[e.g][]{Branch09, Blondin12, Silverman12a,
Silverman12c, Silverman13, Folatelli13} have confirmed this diversity and 
suggest that it could be key to understand the explosion mechanism(s).
In fact, \citet{Maeda10} proposed an explanation in which velocity gradients vary as a
consequence of different viewing directions towards an aspherical explosion scenario.
Nebular  $\ion{[Fe}{ii]}$ $\lambda7155$ and $\ion{[Ni}{ii]}$ $\lambda7378$ \AA\ lines are 
redshifted and are generally associated with HVG SNe, 
while blueshifted lines correspond to LVG SNe. \\
\indent Recent observational evidence suggests the presence of circumstellar material (CSM)  around
SN~Ia progenitors, which in principle could favor the SD model \citep{Raskin13}, but some DD models 
have also presented CSM \citep{Shen13}. In observed spectra, the temporal
evolution in the narrow Na I D lines has been attributed to CSM \citep{Patat07, Simon09, Blondin09},
as well as the fact that they have an excess of blueshifts  \citep{Sternberg11, Maguire13, Phillips13}. \\
\indent It has been suggested that such nearby CSM 
could affect the colors of SNe~Ia \citep{Goobar08,Forster13}, although other studies suggest 
that the dust responsible for the observed reddening of SNe Ia is predominantly located in 
the interstellar medium (ISM) of the host galaxies and not in the CSM associated with the progenitor
system \citep[e.g][]{Phillips13}. \\ 
\indent In this paper we present the optical photometry and optical/near-infrared spectroscopy
of SN~2010ev, a red SN with normal brightness. We discuss its characteristics and we 
compare it with  other similar events. The paper is organized as follows: A description of the 
observations and data reduction are presented in section \ref{s2}. The photometry and spectroscopy are
analyzed in section \ref{s3}. In section \ref{s4} we present the discusion, and in section \ref{s5} 
the conclusions.

\section{Observations and data reduction}
\label{s2}

\begin{figure}
\centering
\includegraphics[width=9cm]{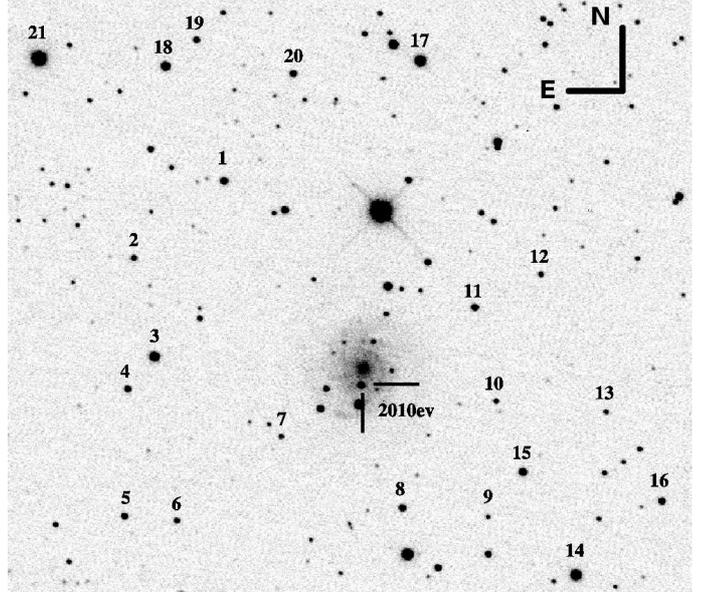}
\caption{Finding chart showing the position of SN~2010ev and that of the local sequence stars used for photometric calibration. 
The image was taken with
PROMPT1 and covers an area of about 8\arcmin $\times$ 7\arcmin. The crosshair indicates the position of the supernova.}
\label{Finfing}
\end{figure}

SN~2010ev was discovered by the Chilean Automatic Supernova Search (CHASE) program on June $27.5$ UT
\citep{Pignata10} in the spiral galaxy NGC 3244 ($\alpha= 10^{\mathrm{h}}25^{\mathrm{m}}28{\fs}99$,
$\delta=-39^\circ49'51{\farcs}2$). The SN lies $1{\farcs}6$ East and $12{\farcs}4$ South of the center
of the host galaxy (see Figure~\ref{Finfing}). Optical spectra of the SN~2010ev were obtained 3 days after
discovery on June $30.9$ UT with the Gemini South (GMOS-S) telescope by \citet{Stritzinger10ev}. The spectrum 
revealed that SN~2010ev was a young ($\sim$ 7 days before maximum) SN~Ia. Details on SN~2010ev and its
host-galaxy properties are summarised in Table~\ref{parameters}.\\

\begin{table}
\caption{Main parameters of SN~2010ev and its host galaxy\label{parameters}}
\centering
\begin{tabular}{l l}
\hline
\hline
Host galaxy			&  NGC 3244\\
Host galaxy type		&  SA(rs)cd $^{\star}$ \\
Redshift			&  0.0092 $^{\star}$ \\
Distance modulus $\mu$		&  32.31$\pm0.60^{\star}$ \\
RA$_{\mathrm{SN}}$		&  $10^{\mathrm{h}}25^{\mathrm{m}}28{\fs}99$ \\
Dec$_{\mathrm{SN}}$		&  $-39^\circ49'51{\farcs}2$ \\
$E(B-V)_{\mathrm{Gal}}$		&  $0.092$ mag $^{\ast}$\\
$E(B-V)_{\mathrm{Host}}$	&  $0.25 \pm 0.05$ mag $^{\dag}$ \\
$\Delta m_{15}(B)$		&  $1.12\pm 0.02^{\bullet}$ \\
Stretch factor(B)		&  $0.94\pm0.01^{\blacktriangle}$ \\
$B_{max}$ epoch (JD)		&  2455384.60$^{\bullet}$ \\
$B_{max}$ epoch (UT)		&  2010 July 7.1 \\
$B_{max}$			&  $14.94\pm0.02^{\bullet}$ \\
$V_{max}$			&  $14.98\pm0.02^{\bullet}$ \\
$V_{max}$ epoch	(JD)		&  2455383.60$^{\bullet}$ \\
$R_{max}$			&  $14.45\pm0.02^{\bullet}$ \\
$R_{max}$ epoch	(JD)		&  2455385.60$^{\bullet}$ \\
$I_{max}$			&  $14.56\pm0.02^{\bullet}$ \\
$I_{max}$ epoch	(JD)		&  2455382.60$^{\bullet}$ \\
$\gamma_B$			&  $1.63\pm0.03^{\triangle}$ \\
$\gamma_V$			&  $1.15\pm0.02^{\triangle}$ \\
$\gamma_R$			&  $1.16\pm0.05^{\triangle}$ \\
$\gamma_I$			&  $0.83\pm0.02^{\triangle}$ \\
\hline
\end{tabular}
\begin{list}{}{}
\item[$^{\star}$] NED (NASA/IPAC Extragalactic Database).
\item[$^{\bullet}$] Obtained with SNooPy.
\item[$^{\ast}$] \citet{Schlegel98}.
\item[$^{\dag}$] See \S \ref{extlc}
\item[$^{\blacktriangle}$] Obtained by SiFTO. 
\item[$^{\triangle}$] Late-time decline $\gamma$ [Magnitudes per 100 days] between 175 and 290 days.
\end{list}
\end{table}

\subsection{Optical photometry}

Optical imaging of SN~2010ev was acquired with the PROMPT1, PROMPT3 and PROMPT5 telescopes located at
Cerro Tololo Interamerican Observatory, FORS2 at the ESO Very Large Telescope (VLT) and IMACS at
Las Campanas Observatory. The PROMPT telescoples are equipped with an Apogee Alta U47 E2V CCD47-10 CCD camera 
(1024$\times$1024, pixel scale = 0.6\arcsec per pixel). With PROMPT1, SN~2010ev was observed with the $B$, $V$, 
$R$ and $I$ Johnson-Kron-Cousins filters, with PROMPT3 it was observed with $B$ filter and the Sloan $u'$, $g'$ 
filters, and in PROMPT5  using $V$, $R$ and $I$ and  $r'$, $i'$ and $z'$ filters. \\
\indent Since the PROMPT cameras operate between -20 and -30 degrees Celsius, all optical images were dark 
subtracted to remove the dark current. After flat-field corrections all images taken with a given 
filter were registered and stacked in order to produce a final deeper image. PSF photometry of the supernova 
was computed relative to a  sequence of stars located close to the SN but not contaminated by host
galaxy light (see Figure~\ref{Finfing}).  The photometric sequence itself was calibrated  to the standard
Johnson Kron-Cousins and Sloan photometric systems using observations of photometric standard stars 
(\citealt{Landolt92}; \citealt{Landolt07}; \citealt{Smith02}), respectively. The  $BVRI$ and
$u'g'r'i'z'$ magnitudes of the local sequence are reported in Table~\ref{stars}. \\
\indent Given that SN~2010ev exploded in a region of significant background galaxy flux, it was necessary
to apply galaxy template subtractions to all of the optical images. Three template images for each filter  
were acquired with the PROMPT telescopes between 2012 January 24--30, i.e. more than 565 days
after $B$ maximum brightness. This makes us confident that the residual SN flux on the template 
images is negligible. Each flux measurement  was computed as a weighted average of the values obtained 
from the three templates. To account for the error introduced by the templates  we add in quadrature 
the rms flux computed from the three measurements with errors obtained from the PSF fitting and flux 
calibration.
In Table~\ref{photom}, we report the $BVRI$ and $u'g'r'i'z'$ photometry of SN~2010ev, 
together with their uncertainties.

\subsection{Optical and near infrared spectroscopy}
 
Optical spectra were obtained at 16 epochs spanning phases between $-6$ and $+270$ days with respect to
$B$-band maximum. These observations were acquired with four different instruments: X-Shooter and FORS2 at 
the ESO Very Large Telescope (VLT), GMOS-S at the Gemini Observatory and the WFCCD at the du Pont Telescope of
the Las Campanas Observatory. Near infrared spectra were obtained with X-Shooter covering 9 epochs from 
$-6$ to $+15$ days. A log of the spectroscopic observations of SN~2010ev is reported in Table~\ref{spectros}. \\
\indent Data reduction for GMOS-S, WFCCD and FORS2 were performed with IRAF\footnote{IRAF is distributed by 
the National Optical Astronomy Observatories (NOAO), which are operated by the Association of   Universities
for Research in Astronomy (AURA), Inc., under cooperative agreement with the National Science Foundation.}
using the standard routines (bias subtraction, flat-field correction, 1D extraction, and wavelength calibration),
while for X-Shooter the dedicated pipeline \citep{Modigliani10} was employed for most of the process, leaving
the telluric line correction and flux calibration to be done with IRAF. To remove the telluric optical and
NIR features, the SN spectrum was divided by the standard star spectrum observed during the same night. The SN 
spectra were flux-calibrated using response curves acquired from the spectra of standard stars.

\begin{table*}[!ht]
\tiny
\centering
\caption{Spectroscopic observations of SN~2010ev.}
\label{spectros}
\begin{tabular}{@ {}l l r c c c l c l l}
\\
\hline \hline
UT date & M.J.D. &           Phase$^{\star}$    &       Range         &   Telescope  &  Arm/Grism$^{\bullet}$  \\
        &        & \multicolumn{1}{r}{[days]}   & \multicolumn{1}{c}{[\AA]}  &  \multicolumn{1}{l}{+ Instrument$^*$} &  &  \\
\hline            
2010/06/30 &  55378.47  &  -6.1    & 3590-9640  &  GEM+GM & B600-500 \& R600-750  \\ 
2010/06/30 &  55378.48  &  -6.1    & 3500-25000 &  VLT+XS & UV/VIS/NIR 		 \\
2010/07/01 &  55379.49  &  -5.1    & 3580-9640  &  GEM+GM & B600-500 \& R600-750  \\
2010/07/03 &  55380.54  &  -4.1    & 3500-25000 &  VLT+XS & UV/VIS/NIR 		 \\
2010/07/04 &  55382.48  &  -2.1    & 3500-25000 &  VLT+XS & UV/VIS/NIR 		 \\
2010/07/05 &  55383.48  &  -1.1    & 3500-25000 &  VLT+XS & UV/VIS/NIR 		 \\
2010/07/06 &  55384.48  &  -0.1    & 3500-25000 &  VLT+XS & UV/VIS/NIR 		 \\
2010/07/07 &  55385.48  &   0.9    & 3500-25000 &  VLT+XS & UV/VIS/NIR 		 \\
2010/07/07 &  55385.49  &   0.9    & 3600-9212  &  DP+WF  & blue       		 \\
2010/07/09 &  55387.49  &   2.9    & 3500-25000 &  VLT+XS & UV/VIS/NIR 		 \\
2010/07/11 &  55389.49  &   4.8    & 3635-9212  &  DP+WF  & blue       		 \\
2010/07/13 &  55391.48  &   6.9    & 3500-25000 &  VLT+XS & UV/VIS/NIR 		 \\
2010/07/21 &  55399.50  &   14.9   & 3500-25000 &  VLT+XS & UV/VIS/NIR 		 \\ 
2010/07/26 &  55404.48  &   19.9   & 3590-8960  &  GEM+GM & B600-500 \& R600-750  \\
2010/12/31 &  55561.49  &   176.9  & 3600-10500 &  VLT+FS & 300V         	 \\
2011/04/03 &  55654.49  &   269.9  & 3600-10500 &  VLT+FS & 300I + OG590 	 \\
\hline \hline
\end{tabular}
\begin{list}{}{}
\item[$^{\star}$] Relative to B$_{max}$ (MJD$=2455384.60)$
\item[$^*$] GEM: Gemini Observatory, GM: GMOS-S, VLT: Very Large Telescope, XS: X-Shooter, DP: Du Pont Telescope, WF: WFCCD, FS: FORS2.
\item[$^{\bullet}$] X-Shooter arm wavelength ranges are UV [3000 - 5600] \AA, VIS [5500 - 12200] \AA, and NIR [10200 - 25000] \AA.
\end{list}
\end{table*}


\section{Results}
\label{s3}

In this section we show the spectral and photometric results obtained for SN~2010ev. The principal 
measurements are compared with other well-studied SNe~Ia that have similar characteristics, such as 
colors, line ratios and velocities. In order to interpret our observations and results, we
compare them with the Hsiao SN~Ia spectral template \citep[][hereafter ``H07'']{Hsiao07} and
synthetic spectra computed from a delayed-detonation model \citep[][hereafter ``B15''
\footnote{Synthetic spectra obtained from: 
https://www-n.oca.eu/supernova/snia/sn2002bo.html}]{Blondin15}. 
\citet{Hsiao07} use a sample of 28 SNe~Ia to characterize the spectral
features and identify patterns in the data with principal component
analysis (PCA). Overall, these SNe show normal features. Meanwhile, the B15 model is the result of a 
1D non-local thermodynamic equilibrium radiative transfer simulation of a Chandrasekhar mass delayed-detonation
model with 0.51$M_{\odot}$ of $^{56}$Ni that closely matches SN~2002bo. This model provides a reference for understanding SNe Ia similar to that
prototype, as is the case for SN~2010ev. Thus, we compare our results with the H07 template and the B15 model.

\subsection{Light curves}

\begin{figure}[hbtp]
\centering
\includegraphics[width=\columnwidth]{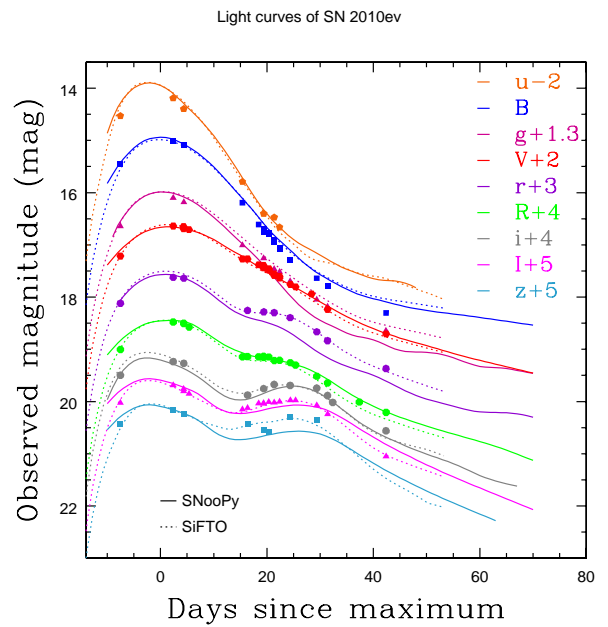}
\caption{$BVRI$ and $u'g'r'i'z'$ light curves of SN~2010ev. The SNooPy fits are shown in solid lines while
the SiFTO fits in dotted lines. The light curves have been shifted by the amount shown in the label.}
\label{cl}
\end{figure}

SN~2010ev was observed in $BVRI$ and $u'g'r'i'z'$ bands. We have performed light curve fits to 
the multi-wavelength photometry of SN~2010ev. For this purpose, we use SNooPy \citep{Burns11} and
SiFTO \citep{Conley08} light curve fitters. Figure~\ref{cl} shows the $BVRI$ and $u'g'r'i'z'$ light
curves with both fits. This SN shows a normal decline rate, 
$\Delta m_{15}(B)=1.12 \pm 0.02$ and a stretch parameter $s=0.94\pm 0.01$. 
This $\Delta m_{15}(B)$ is similar to those found in high velocity gradient SNe (HVG), 
such as SN~2002bo ($\Delta m_{15}(B)=1.13 \pm 0.02$)
and SN~2002dj ($\Delta m_{15}(B)=1.08 \pm 0.02$).

\indent Using SNooPy we obtain a peak $B$-band magnitude $B_{max}=14.94\pm0.02$ 
on $JD=2455384.60\pm0.30$ (2010 July 7.1 UT),
which indicates that SN 2010ev was observed in $BVRI$ and $u'g'r'i'z'$ from $-7.5$ to $289.5$ days 
with respect to maximum light. The peak $VRI$ magnitudes are 
$V_{max}=14.98\pm0.02$, $R_{max}=14.45\pm0.02$ and $I_{max}=14.56\pm0.01$, that occur 
at $-1$, 1 and $-2$ days with respect to B$_{max}$. The $I$ and $i'$ bands show a
secondary maximum at $\sim20-25$ days after B maximum, while $R$ and $r'$ bands show a shoulder 
at those times. The main photometric parameters of SN~2010ev are reported in Table~\ref{parameters}.\\
\indent During the nebular phase the $BVRI$ magnitudes follow a linear decline
due to the exponentially decreasing rate of energy input by radioactive decay: $1.63\pm0.03$, $1.15\pm0.02$,
$1.16\pm0.05$, $0.83\pm0.02$ magnitudes per 100 days, respectively. The slope of the $B$ light curve is higher
that those found by \citet{Lair06} but lower that those in the $V$, $R$ and $I$ bands. 
Despite these differences, these decline rates are consistent with other well studied SNe~Ia
(e.g., \citealt{Stanishev07, Leloudas09}), which show the same slower decline in the $I$ band.

\subsection{Color Curves}

\begin{figure}[hbtp]
\centering
\includegraphics[width=\columnwidth]{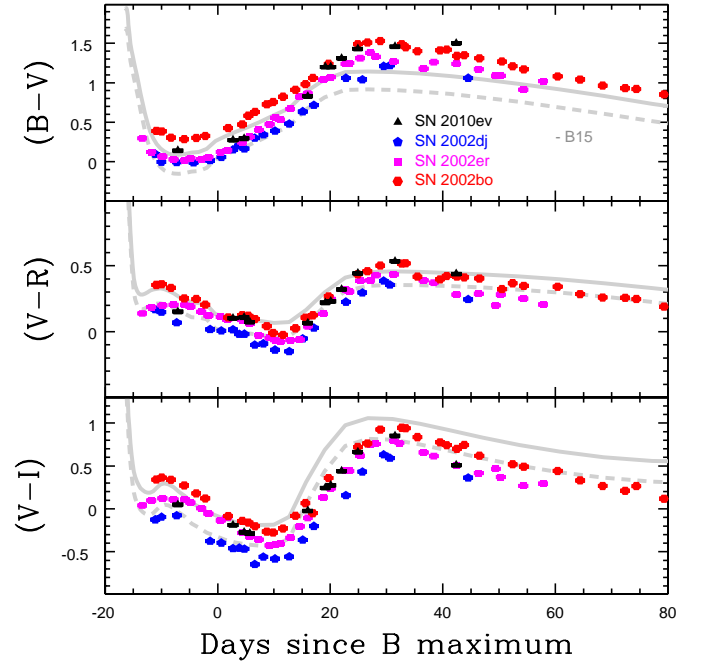}
\caption{Color evolution of SN~2010ev compared with high velocity gradient (HVG) SNe~Ia: SN~2002bo, 
SN~2002dj and SN~2002er. The SN colors have been dereddened for MW extinction only. We also show the
colors of the B15 model without extinction (solid grey line) and with the host extinction (dashed grey line) obtained in section~\ref{extlc}.}
\label{colevol}
\end{figure}

The $(B-V)$, $(V-R)$ and $(V-I)$ color curves of SN~2010ev are compared in Figure~\ref{colevol} with 
SN~2002bo  \citep{Benetti04}, SN~2002dj \citep{Pignata08} and SN~2002er \citep{Pignata04}, 
as well as the delayed-detonation B15 model (grey lines) for SN~2002bo.
The colors have been corrected for Milky Way (MW) extinction exclusively. 
The B15 model colors were obtained with synthetic photometry by integrating the model 
spectral energy distributions (SEDs).\\
\indent  At maximum, these SNe all have redder $B-V$ colors than the typical average SN~Ia color 
($B-V\sim0$), as represented by the B15 model. 
Before maximum, SN~2002bo has redder $(B-V)$ colors than SN~2010ev, but 
around 20 days they have similar colors. Meanwhile,
SN~2002dj and SN~2002er are bluer at all phases. The B15
model is bluer in $B-V$ than all SNe at all epochs indicating the high reddening 
in the line of sight within the host galaxies of these SNe. This is true when using the H07 template as well.\\
\indent The peak of the $(B-V)$ color evolution happens around 30 days, compared to around 26 days
in the B15 model.
This evolution is similar in $(V-R)$ and $(V-I)$. 
The difference in the time of $(B-V$) maximum and the color evolution have shown to be very important 
for SNe~Ia \citep{Burns14,Forster13}. According to \citet{Blondin15} a shift of 5 days earlier/later
in the  $(B-V$) maximum correspond to a decrease/increase of 0.1$M_{\sun}$ of $^{56}$Ni synthesized 
during the explosion. Since these SNe have similar peak bolometric luminosity (see \S~\ref{bolom}),
the diversity seen in Figure~\ref{colevol} could be attributed to only small changes in $^{56}$Ni mass,
which also affect the temperature and ionization.

\subsection{Optical spectral evolution}

\subsubsection{Early phases}

\begin{figure*}[hbtp]
\centering
\includegraphics[width=18cm]{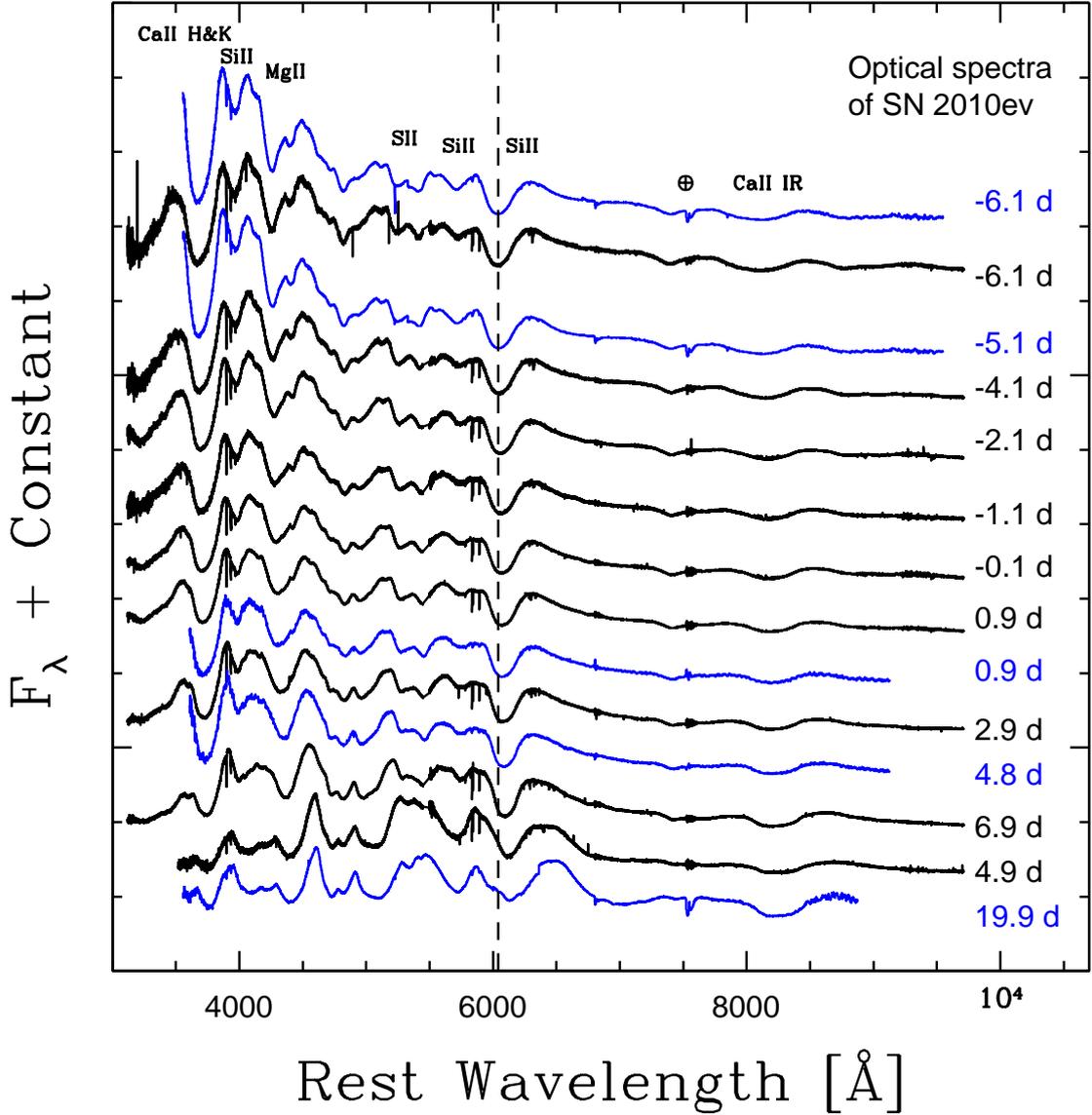}
\caption{Spectroscopic sequence of SN~2010ev ranging from $-6.1$ to 19.9 days around B-band maximum.
Each spectrum has been corrected for Milky Way reddening and shifted by an arbitrary amount for presentation.
We show low resolution spectra in blue and medium resolution spectra in black.
The phases are labeled on the right.}
\label{spectravis}
\end{figure*}

Figure~\ref{spectravis} shows the optical spectra evolution of SN~2010ev from -6.1 to 19.9 days. The spectra
show that SN~2010ev is a normal SN~Ia with very prominent \ion{Si}{ii} $\lambda6355$ \AA\ absorption.
Pre-maximum spectra exhibit characteristic P-Cygni profiles of \ion{Si}{ii} $\lambda4130$, $\lambda5972$ and
$\lambda6355$; \ion{Ca}{ii} H \& K $\lambda3945$ and IR triplet $\lambda8579$; \ion{S}{ii} $\lambda5449$ and
$\lambda5622$ \AA. Other lines such as $\ion{Mg}{ii}$ $\lambda4481$ \AA, and some blends caused by $\ion{Fe}{ii}$ 
in the 4500 to 5500 \AA\ range are clearly visible. Despite contamination from the telluric feature near
$\lambda7600$, \ion{O}{i} $\lambda7774$ \AA\ is also detected. 
The narrow $\textrm{Na\,{\sc i}\,D}$ and \ion{Ca}{ii} H \& K from the host galaxy and
the MW, as well as \textit{diffuse interstellar bands} (DIBs) at $\lambda 5780$ and $\lambda 6283$ \AA\
are also present, which suggest significant reddening.

\begin{figure}[hbtp]
\centering
\includegraphics[width=\columnwidth]{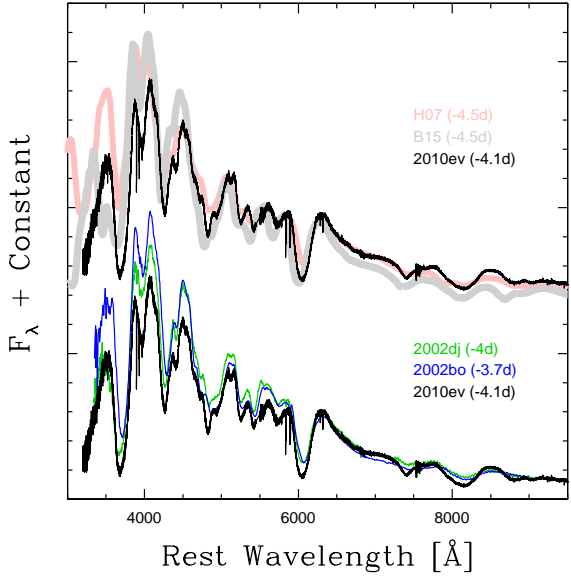}
\caption{Comparison of pre-maximum (around $-4$ days) spectra of SN~2010ev, SN~2002bo, SN~2002dj,
the H07 template and B15 model. 
The spectra have been corrected by MW reddening and redshift. Epochs are marked in the plot.} 
\label{comparacion_4}
\end{figure} 

\begin{figure}[hbtp]
\centering
\includegraphics[width=\columnwidth]{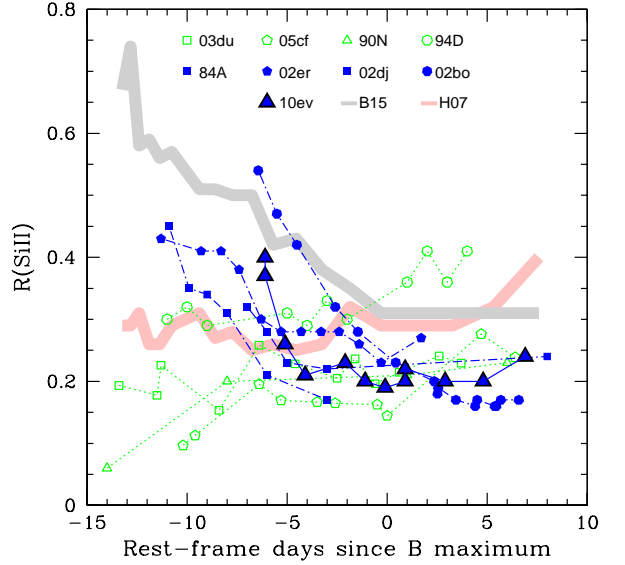}
\caption{Evolution of ${\cal R}$(\ion{Si}{ii}) for SN~2010ev compared with a sample of HVG SNe (blue
filled symbols) and LVG SNe (empty green symbols). In grey is shown the evolution of ${\cal R}$(\ion{Si}{ii})
for the B15 model and in pink for H07 template.} 
\label{ratiosi}
\end{figure}

\indent In Figure~\ref{comparacion_4} the optical spectrum of SN~2010ev at approximately $-4$
days from $B$-band maximum is compared 
at the same epoch with SNe with very prominent \ion{Si}{ii} $\lambda6355$ \AA\ absorption and similar
colors, such as SN~2002bo and SN~2002dj. The H07 template and B15 model are also shown for
comparison. 
As can be seen, SN~2010ev shows stronger \ion{Si}{ii} $\lambda6355$
absorption features compared with SN~2002bo and SN~2002dj, and similarities in lines like 
\ion{Ca}{ii} and \ion{S}{ii}. Since the B15 model is well matched with SN~2002bo, their lines widths and
the pseudo-continuum are very similar, while the H07 template shows smaller absorption lines of \ion{Si}{ii} 
$\lambda6355$ \AA\ and the \ion{Ca}{ii} IR triplet. The \ion{O}{i} $\lambda7774$ \AA\ line is more prominent 
in SN~2010ev than the other SNe, which could suggest either 
differences in the amount of unburned material or in the oxygen abundance, produced by C burning.
Considering its velocity ($\sim14500$ km s$^{-1}$), it could be attributed to unburnt C
\citep{Blondin15}. However, we can not confirm the latter using the possible presence of
\ion{C}{ii} due to a lack of very early spectra. \\
\indent  At maximum, the ratio of the depth of the \ion{Si}{ii} 
$\lambda5972$ and $\lambda6355$ absorption features, ${\cal R}$(\ion{Si}{ii}) \citep{Nugent95} is 
${\cal R}$(\ion{Si}{ii})$\,=0.20\pm0.03$, while the pseudo-equivalent widths (pEWs) give $150.80\pm1.21$ \AA\ and
$15.91\pm0.72$ \AA\, respectively. Based on the strength of the \ion{Si}{ii} lines defined by \citet{Branch06}, 
SN~2010ev is a Broad-Line (BL) SN. 
The evolution of ${\cal R}$(\ion{Si}{ii}) of SN~2010ev is compared in Figure~\ref{ratiosi} with HVG and low 
velocity gradient (LVG, \citealt{Benetti05}) SNe. As can be seen, SN~2010ev shows a dramatic decline before 
maximum from ${\cal R}$(\ion{Si}{ii})$=0.40$ at $-6$ days to ${\cal R}$(\ion{Si}{ii})$=0.20$ around maximum.
Then, it shows a flat evolution, which is consistent with HVG SNe. This behavior reflects  
lower temperatures before maximum in the spectrum-forming region, which then increase. 
Figure~\ref{ratiosi} also shows the evolution of ${\cal R}$(\ion{Si}{ii}) for H07 template and 
B15 model. The B15 model is consistent with the evolution of the HVG SN~2002bo;
meanwhile, the evolution of H07 template shows a behavior similar to LVG SNe.
	
After maximum, the \ion{Ca}{ii} IR triplet (Figure~\ref{spectravis}) becomes very prominent, while the
\ion{Si}{ii} $\lambda5972$ and \ion{S}{ii} lines fade rapidly.
The \ion{S}{ii} lines are not detectable $\sim2$ weeks after maximum whereas \ion{Si}{ii} $\lambda6355$ is visible 
for $\sim20$ days. At 14 days after maximum the \ion{O}{i} $\lambda7774$ line disappears and \ion{Ca}{ii} H \& K decreases 
significantly. At around 20 days lines from iron-group elements start to dominate the spectrum,
as the SN ejecta layers expand and become more transparent. \\

\subsubsection{Late phases}

\begin{figure*}[hbtp]
\centering
\includegraphics[width=13.8cm, height=12cm]{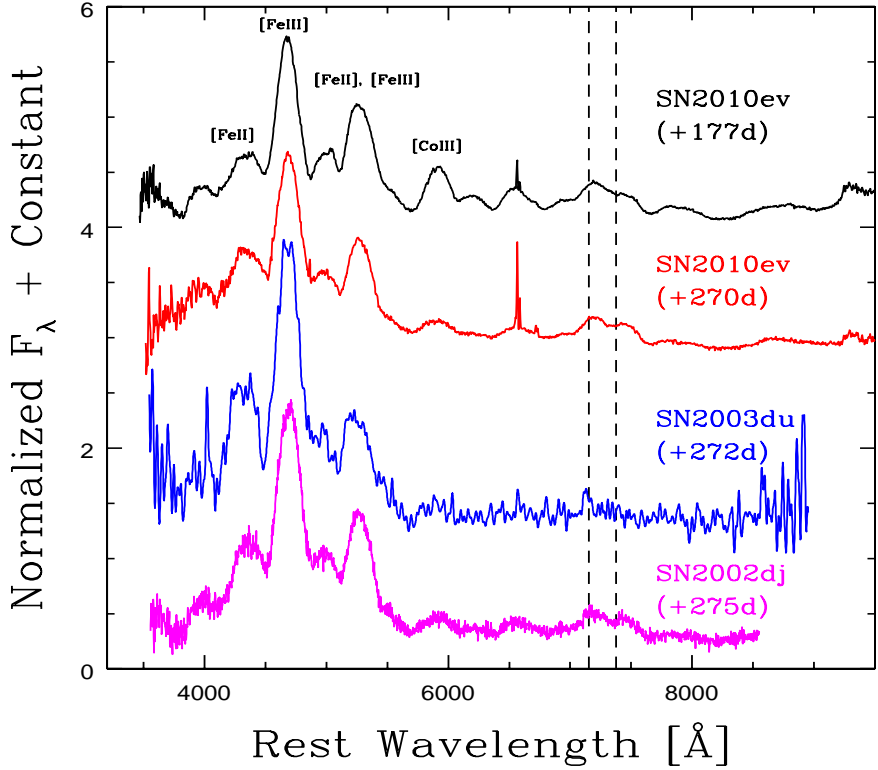}
\caption{Nebular spectra of SN~2010ev taken at 177 and 270 days compared with SN~2003du, and SN~2002dj around 270 days.
The spectra have been corrected for redshift and normalized with respect to the SN~2010ev 
flux in $V$-band (and shifted by an arbitrary constant).The main features have been labeled, while the epochs 
and the SN name are shown on the right. The dashed lines are the rest position of \ion{[Fe}{ii]} $\lambda7155$ and
\ion{[Ni}{ii]} $\lambda7378$}
\label{neb}
\end{figure*}

\indent In the nebular phase, two spectra were obtained at $\sim$177 and $\sim$270 days with FORS2.
In this phase, the spectrum is mainly dominated by forbidden lines of iron-group elements: 
\ion{[Fe}{ii]}, \ion{[Fe}{iii]}, \ion{[Ni}{iii]}, \ion{[Ni}{iii]} and \ion{[Co}{ii]}, which were 
identified in SN~2010ev (see Figure~\ref{neb}). The spectra also show typical lines of an
\ion{H}{ii} region at the SN site such as H$_{\alpha}$, \ion{[N}{ii]}, and \ion{[S}{ii]}.
The strongest feature at this epoch is the blend of \ion{[Fe}{iii]} 
lines at $\lambda4701$ \AA\ \citep{Maeda10a}. The velocity offset of peak emission shows a significant 
temporal change from $1300\pm100$ km s$^{-1}$ at 177 days to $490\pm20$ km s$^{-1}$ at 270 days from
the rest position. This behavior is consistent with that found by \citet{Maeda10a} for a sample
of 20 SNe~Ia with late-time nebular spectra and different velocities, light-curve widths and colors.
Meanwhile, the FWHM velocities show the opposite trend: At 177 days, 
the FWHM=$14800\pm300$ km s$^{-1}$ and increases to  $16400\pm600$ km s$^{-1}$ at 270 days. 
Taking an average of the relation derived by \citet{Mazzali98} and more recently by \citet{Blondin12}, 
we can infer $\Delta m_{15}(B)=1.10\pm0.03$ based on the FWHM velocities of $\ion{[Fe}{iii]}$ at $t>200d$,
which is consistent with the one obtained with SNooPy. However, it should be noted that this relation is not significant 
when subluminous events are excluded \citep{Blondin12, Silverman13}.

\indent Other lines in the spectra seem to have no significant evolution, except the emission lines 
near $\sim6000$ \AA, which appear to decrease with time, and the blend of \ion{[Fe}{ii]}
$\lambda7155$ and \ion{[Ni}{ii]} $\lambda7378$ features that develop a double-peaked profile. \\
\indent In Figure~\ref{neb} the nebular spectra of SN~2010ev are compared with SN~2003du \citep{Stanishev07}
and SN~2002dj \citep{Pignata08} around 270 days. The $\sim4700$ \AA\ feature is similar in SN~2010ev and SN~2002dj, 
although slightly more pronounced in the latter. In SN~2003du this feature appears to be stronger.
Also, the \ion{[Fe}{ii]} $\lambda7155$ and \ion{[Ni}{ii]} $\lambda7378$ lines are blueshifted.
This shift may suggest an asymmetry during the initial deflagration of the explosion in the direction
away from the observer \citep{Maeda10}. At 270 days, we find  $v_{neb}=2150\pm220$ km s$^{-1}$,
inferred from the average of the Doppler shifts of the emission lines of $\ion{[Fe}{ii]}$ $\lambda7155$ and 
$\ion{[Ni}{ii]}$ $\lambda7378$. Redshifted 
nebular velocities have been seen to relate with HVG and redder colors  \citep{Maeda11,Forster12}
and with narrow \ion{Na}{i} D equivalent width \citep{Forster12}. We confirm these trends with SN~2010ev.

\subsection{NIR Spectral evolution}

\begin{figure*}[hbtp]
\centering
\includegraphics[width=17cm]{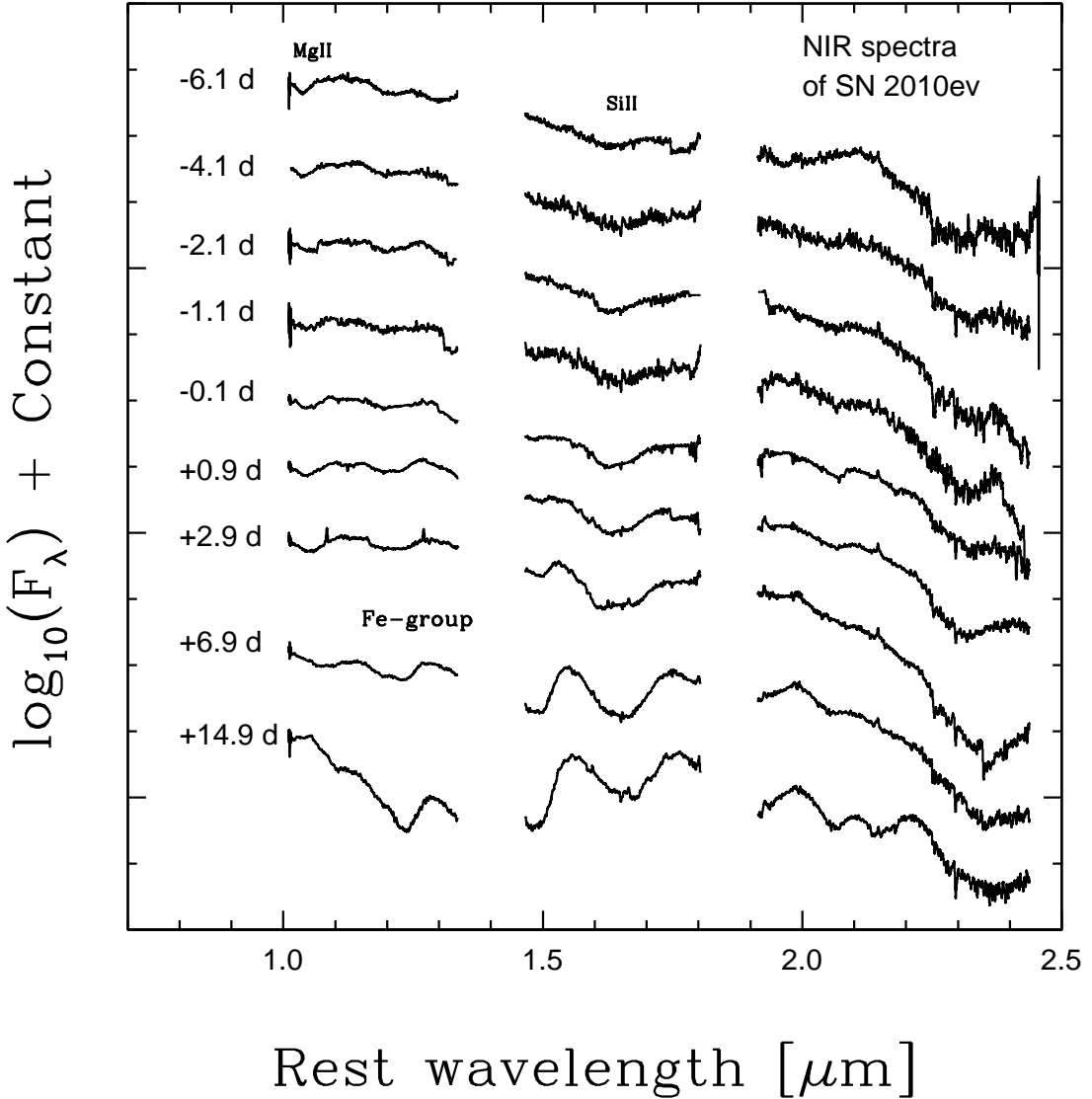}
\caption{NIR Spectra of SN~2010ev taken between $\sim-7$  and $\sim15$ days with X-Shooter.
The spectra are displayed in log scale. Each spectrum have been corrected for redshift and shifted by
an arbitrary amount for presentation. The phases are labeled on the left.} 
\label{spectra_ir}
\end{figure*}

The NIR spectra of SN~2010ev between $-6$ to 15 days with respect to B$_{max}$ are presented in 
Figure~\ref{spectra_ir}. The early spectra show a blue pseudo-continuum with a weak feature at $\sim$10500 \AA\,
which corresponds to \ion{Mg}{ii} $\lambda10927$ \citep{Wheeler98}. The strength of this 
feature seems to be constant with time, while other lines are getting stronger. Near $\sim16500$ \AA\   
a weak feature
is clearly visible, which has been identified as \ion{Si}{ii} by \citet{Gall12} and as \ion{Fe}{iii} by 
\citet{Hsiao13}. Near $\sim20800$ \AA\ we detect a feature which has not been clearly identified,
but according to \citet{Benetti04} this line is due to \ion{Si}{ii}, while \citet{Stanishev07} 
suggest that the line is \ion{Si}{iii}. \ion{C}{i} $\lambda10693$ is not detected in our spectra,
but possibly contributes to \ion{Mg}{ii} $\lambda10927$.

\indent The $H$-band break ratio ($R=f_1/f_2$) defined by \citet{Hsiao13} as the ratio
between the maximum flux level redwards of 1.5 $\mu$m ($f_1$) and the maximum flux just bluewards
of 1.5 $\mu$m ($f_2$), can be seen in the spectra
of SN~2010ev at 2.9 days. The break at this epoch increases from $R=1.26\pm0.14$ to $2.14\pm0.11$ 
at 6.9 days and takes the maximum value at 14.9 days ($R=3.11\pm0.09$). \citet{Hsiao13}
found that this parameter appears to peak uniformly around 12 days past $B$-band maximum, and that 
it is correlated with $\Delta m_{15}(B)$. Using the mean decline rate estimated by \citet{Hsiao13}
for a sample of SNe~Ia, we measure the ratio at 12 days and find R$_{12}=3.39\pm0.15$, 
which corresponds well with our $\Delta m_{15}(B)$ estimate (\citealt{Hsiao13}, Figure~11) .\\
\indent At 6.9 days the spectrum shows emission features present at 15500 \AA\ and 17500 \AA. These 
features are attributed to blends of iron goup elements: \ion{Co}{ii}, \ion{Fe}{ii} and \ion{Ni}{ii}
(\citealt{Wheeler98}; \citealt{Marion03}). Above 20000 \AA, lines of \ion{Co}{ii}  dominate the spectrum 
\citep{Marion09}. The presence of these lines means that the spectrum-forming region has receded enough
to reach the iron group dominated region.\\

\subsection{Expansion velocities}

\begin{figure}[hbtp]
\centering
\includegraphics[width=\columnwidth]{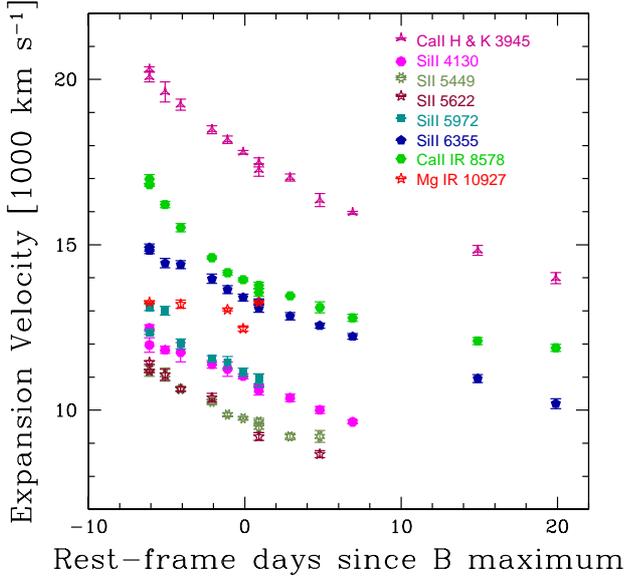}
\caption{Evolution of expansion velocities of SN~2010ev derived from the 
maximum absorptions of different lines.}
\label{velocidad}
\end{figure}

\begin{figure}
\centering
\includegraphics[width=\columnwidth]{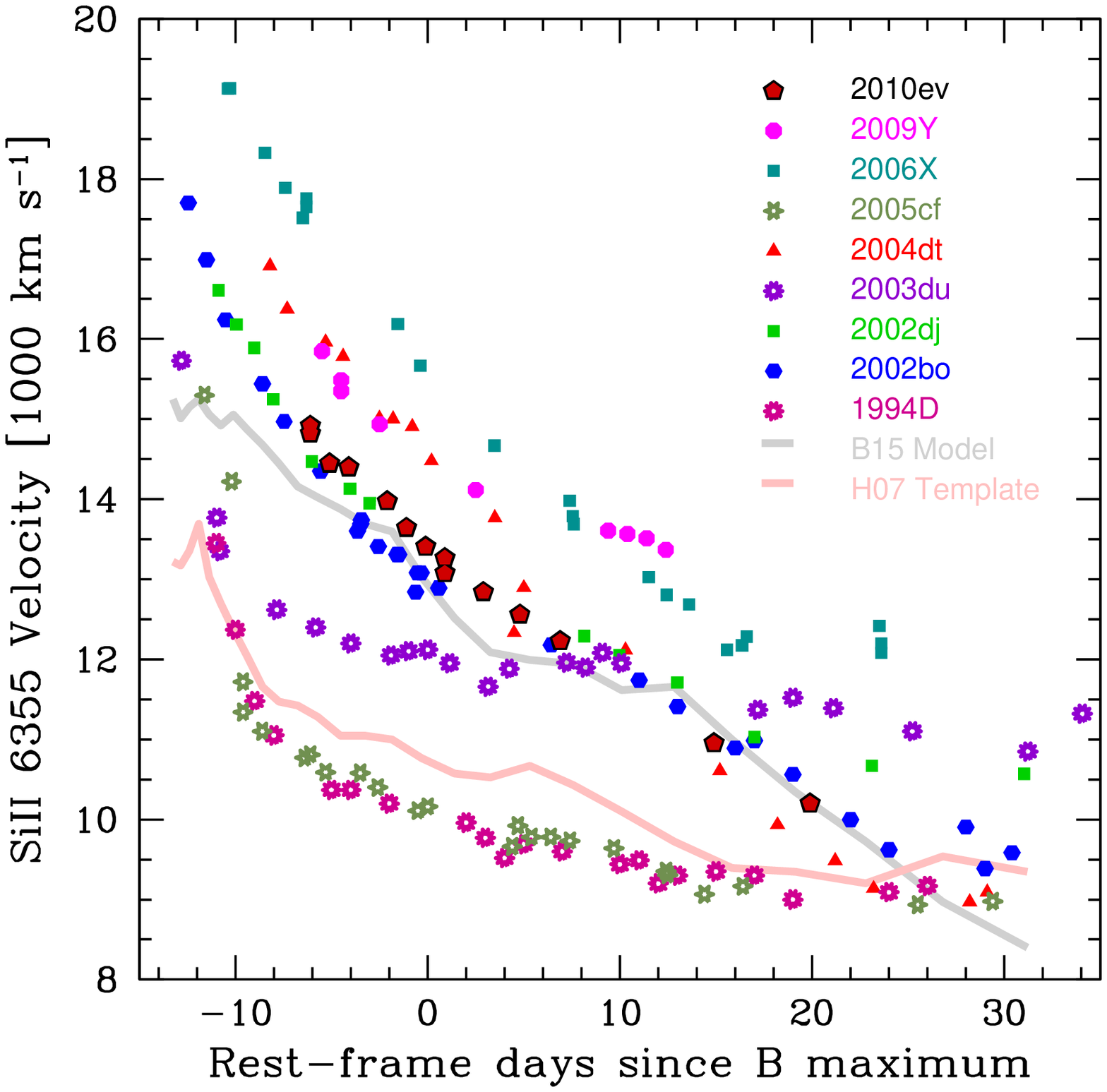}
\caption{$\ion{Si}{ii}$ $\lambda6355$ expansion velocity evolution
of SN~2010ev derived from the minimum of the absorption line, compared with 
other SNe: SN~2009Y, SN2006X \citep{Wang08}, SN2005cf \citep{Pastorello07}, 
SN2004dt \citep{Altavilla07}, SN2003du \citep{Stanishev07}, SN2002dj \citep{Pignata08},
SN2002bo \citep{Benetti04}, SN1994D \citep{Patat96}, H07 template, and B15 model.}
\label{compvel}
\end{figure}

The analysis of the spectra indicate large and rapidly decreasing expansion velocities due to 
the rapidly receding spectrum-forming region to deeper layers with time. In Figure~\ref{velocidad},
we present the velocity evolution for selected lines of
\ion{Si}{ii}, \ion{Ca}{ii}, \ion{S}{ii} and \ion{Mg}{ii}. It clearly 
shows that the expansion velocity of \ion{Ca}{ii} is higher than \ion{Si}{ii}. The \ion{Si}{ii} 
minimum evolves from 14800 km s$^{-1}$ at -7 days to 10200 km s$^{-1}$ at 19 days, while at the same 
epoch \ion{Ca}{ii}  H \& K decrease from 20100 to 14000 km s$^{-1}$ and the \ion{Ca}{ii} IR triplet
from 17000 to 11900 km s$^{-1}$. This implies that the \ion{Ca}{ii} lines mostly form in the outer
shell of the ejecta, while \ion{S}{ii}, which has a higher ionization potential, forms in deeper
layers, resulting in lower absorption velocities 
(11400 at -7 days to 8600 km s$^{-1}$ at 5 days and then disappears). Meanwhile, 
\ion{Mg}{ii} $\lambda$10900 \AA\ shows a nearly constant velocity, which is consistent with 
the findings of \citet{Hsiao13}, who show that the velocity is remarkably constant after 
a short period of decline in very early phases. After 1 day past maximum, the \ion{Mg}{ii} feature
is difficult to measure due to the blend with other lines. \\
\indent From the velocity evolution of $\ion{Si}{ii}$ $\lambda6355$ between maximum and 20 days,
we obtain a velocity gradient of $\dot v_{\mathrm{Si}}=164\pm7$ km s$^{-1}$ d$^{-1}$, which places
SN~2010ev among the HVG group \citep{Benetti05}. This result is comparable 
with the definitions of velocity gradient put forward by \citet{Blondin12} and \citet{Folatelli13}. 
In the former we obtain $\Delta v_{abs}/\Delta t_{[+0,+10]}=166\pm14$ km s$^{-1}$ d$^{-1}$, while 
in the latter we find $\Delta v_{20}(\mathrm{Si})=3210\pm183$ km s$^{-1}$. To be consistent with the units,
we divide this last value by 20 days and we obtain $160.5\pm9.2$ km s$^{-1}$ d$^{-1}$. 
Since the \ion{Si}{ii} velocity in SN~2010ev is quasi-linear, all three gradients agree 
with each other.\\
\indent In Figure~\ref{compvel} we compare the time evolution of the expansion velocity of 
$\ion{Si}{ii}$ $\lambda6355$ with eight well studied SNe~Ia. It can be clearly seen that the velocity 
evolution of SN~2010ev, SN~2002bo and SN~2002dj are consistent with the HVG class. In contrast,
SN~1994D and SN~2005cf belong to the LVG group. Table~\ref{veldecl} shows
the velocity gradient for these SNe measured in different ways. SN~2010ev has one of the highest $\dot v_{\mathrm{Si}}$ value.
Figure~\ref{compvel} also shows the velocity evolution for H07 template and B15 model. As noted above,
the B15 model gives better agreement with our SN, while the H07 templete gives better results
for the LVG group.  

\begin{table}[!ht]
\centering
\caption{Velocity decline for the sample used in Figure~\ref{compvel}.
The second column is the mean velocity decline between maximum and +10 days \citep{Blondin12}. 
The third column is estimated in the same way but between maximum and 20 days 
\citep{Folatelli13}. The last column is derived doing a fit between maximum and the last available value \citep{Benetti05}.}
\label{veldecl}
\begin{tabular}{l c c c}
\hline
\hline
SN 	& $\Delta v_{abs}/ \Delta t_{[+0,+10]}$ &  $\Delta v_{20}(Si)$/20	& $\dot v_{\mathrm{Si}}$ $^{\star}$\\
	&	[km s$^{-1}$ d$^{-1}$]		&  [km s$^{-1}$ d$^{-1}$]	& [km s$^{-1}$ d$^{-1}$ ] \\
\hline
2003du	&  $17$					&  $33$  			&  $31$	 		\\
2005cf	&  $52$					&  $54$				&  $35$ 		\\
1994D 	&  $64$ 				&  $54$				&  $39$			\\
2009Y	&  $96$					&  $86$				&  $125$  		\\
2002bo	&  $122$				&  $115$			&  $110$  		\\
2002dj	&  $145$				&  $132$ 			&  $86$ 		\\
2010ev	&  $166$				&  $160$			&  $164$		\\
2006X	&  $235$				&  $179$			&  $123$		\\
2004dt	&  $244$				&  $245$			&  $160$   		\\
\hline
\end{tabular}
\begin{list}{}{}
\item [$\star$] Taken from \citet{Maeda10}, except the value of SN~2010ev, which was estimated in this work.  
\end{list}
\end{table}

\subsection{Extinction from the light-curve}\label{extlc}

The nature of red colors towards SN~Ia is still debated. It is not clear what is intrinsic to the SN 
and what is due to reddening from material in the line of sight. Recent claims
of circumstellar interaction have fed the question of whether their color evolution and the atypical 
inferred host extinction laws actually relate to nearby material ejected close to explosion.
In this section we explore different methods to estimate the reddening and extinction law towards 
SN~2010ev, as well as any other evidences for CSM from a photometric perspective. The $B-V$ color at 
B$_{max}$ obtained from SiFTO is $0.29\pm0.06$. This value is above the typical values of the type Ia 
SNe which have color at maximum between $-0.2<B-V<0.2$ \citep[e.g][]{Gonzalez14}, so that the host-galaxy 
extinction appears to be significant from a photometric point of view.
With the relation proposed by \citet{Phillips99}, using the
maximum-light colors we estimate $E(B-V)_{\mathrm{Host}}=0.26\pm0.07$. This result is consistent with
the value obtained through the relation of \citet{Folatelli10}: $E(B-V)_{\mathrm{Host}}=0.29\pm0.05$
and with $E(B-V)_{\mathrm{Host}}=0.29\pm0.02$ given by SNooPy. We summarize these findings in 
Table~\ref{ebv-table}.

\begin{table*}[!ht]
\centering
\caption{Line of sight extinction $A_V$, reddening law $R_V$ and color excess $E(B-V)$ for SN~2010ev
according to different spectroscopic and photometric techniques. }
\label{ebv-table}
\begin{threeparttable}
\renewcommand{\arraystretch}{1.15}
\begin{tabular}{c c c l}
\hline
\hline
$A_{V}$ &  $R_V$ & $E(B-V)$ & Reference \\
\hline
\multicolumn{4}{c}{\textbf{MILKY WAY}}\\

$0.28\pm0.06$ $^{\star}$  & $\cdots$              & $\cdots$              & MW dust extinction maps \citep{Schlafly11} \\
$\cdots$                   & $\cdots$              & $0.147\pm0.003$       & EW(\ion{Na}{i} D) via \citet{Turatto03}	\\
$\cdots$                   & $\cdots$              & $0.169\pm0.034$       & EW(\ion{Na}{i} D) via \citet{Poznanski12}\\
$0.28\pm0.02$             & $\cdots$              & $\cdots$              & MW \ion{Na}{i} D column density \citep{Phillips13} \\
\hline
\multicolumn{4}{c}{\textbf{HOST}}\\
$\cdots$                   & $\cdots$              & $0.26\pm0.07$         & Maximum light colors via \citet{Phillips99}   \\
$\cdots$                   & $\cdots$              & $0.29\pm0.05$         & Maximum light colors via \citet{Folatelli10}  \\
$\cdots$                   & $\cdots$              & $0.29\pm0.02$         & SNooPy fit \citep{Burns11}	\\
$0.50^{+0.17}_{-0.19}$     & $1.54^{+0.57}_{-0.59}$& $\cdots$              & MCMC light-curve fit \citep{Phillips13,Burns14} \\
--                         & $1.54\pm0.65$         & $0.25\pm0.05$         & Color excess fit (this work) \\
\hdashline
$\cdots$                   & $\cdots$              & $0.107\pm0.008$       & EW(\ion{Na}{i} D) via \citep{Turatto03}	\\
$\cdots$                   & $\cdots$              & $0.085\pm0.050$       & EW(\ion{Na}{i} D) via \citep{Poznanski12}	\\
$0.38\pm0.02$              & $\cdots$              & $\cdots$              & \ion{Na}{i} D column density \citep{Phillips13} \\
$\cdots$                   & $\cdots$              & $0.53\pm0.09$         & EW(DIB) $\lambda$5780 \AA\ via \citep{Luna08} \\
$1.18\pm0.01$              & $\cdots$              & $\cdots$              & EW(DIB) $\lambda$5780 \AA\ \citep{Phillips13} \\
$\cdots$                   & $\cdots$              & $0.50\pm0.04$         & EW(DIB) $\lambda$6283 \AA\ via \citep{Luna08} \\
$0.24\pm0.03$              & $\cdots$              & $\cdots$              & \ion{K}{i} column density \citep{Phillips13} \\
\hdashline
$\cdots$                   & $\lesssim$ 2          & $\cdots$              & Continuum polarization \citep{Zelaya15} \\

\hline
\end{tabular}
\begin{list}{}{}
\item [$^{\star}$] The error is calculated from the difference with \citet{Schlegel98}.
\end{list}
\end{threeparttable}
\end{table*}

\indent Extensive evidence \citep[e.g][]{Riess96,Elias-Rosa06,Conley07,Krisciunas07,Goobar14} 
suggests that at least some SNe~Ia suffer 
from a lower characteristic $R_V$ reddening law than the Galactic average value of $R_V=3.1$ 
\citep{Fitzpatrick07}. It has been claimed that such variation could be attributed to CSM near
the supernova \citep{Wang05,Goobar08,Amanullah11}. In fact, there is an intriguing trend of low $R_V$'s
and high extinction towards SNe \citep{Mandel11,Kawabata14} which raises the question of whether interstellar extinction
towards extragalactic sites with large amounts of dust is different from the  Milky Way (MW), 
or if some nearby material affects the color of SNe~Ia in such a way as to mimic this effect. SN~2010ev 
is reddened and is thus a good candidate for low $R_V$. \\
\indent In order to estimate a reddening law for SN~2010ev, we calculate the color excesses at maximum at 
different wavelengths to fit them to various reddening laws in a similar way to \citet{Folatelli10}.
Firstly, we obtain  colors $(V-X)$ at $B$-band maximum light for bands $X=u',B,g',r'$ and $i'$ obtained 
from our SiFTO fit. These colors have been $K$-corrected through the H07 template warped  
to the observed photometric colors, and then corrected for MW extinction. To obtain color excesses 
we use intrinsic colors from both the H07 template and the B15 model.

\begin{figure}[hbtp]
\centering
\includegraphics[width=9cm]{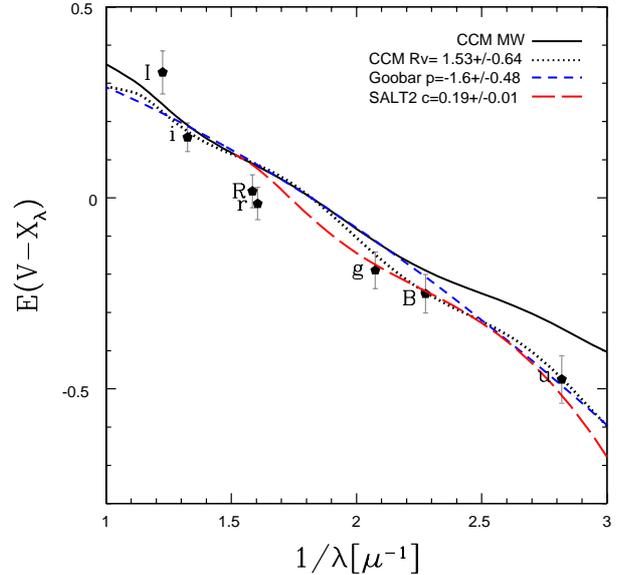}
\caption{Color excesses $E(V-X)$ vs $1/\lambda$ for SN~2010ev.  
Lines are fits to the excesses with a standard  $R_v=3.1$ (solid black) and  free  $R_v=1.54$ (black dotted)
\citet{Cardelli89} extinction law, a \citet{Goobar08} law (blue dashed) and a SALT2 
color law \citep{Guy07} fit (dotted dashed red).}
\label{excess-fit}
\end{figure}

\indent The resulting color excesses using intrinsic colors from the B15 model are shown in Figure~\ref{excess-fit}, 
where we also show different 
reddening law fits. The best reddening law we find for \citet{Cardelli89}, modified by \citet{ODonnell94} (CCM)   
is $R_v=1.54\pm0.65$ with $E(B-V)=0.25\pm0.05$, which is consistent with the model by \citet{Fitzpatrick99}
($R_v=1.72\pm0.60$), and is also consistent with the reddening law of \citet{Goobar08} for circumstellar dust.
The reddening law of SN~2010ev is different from standard values for the MW and is consistent with other values
of reddened SNe. This argues for different dust properties such as size in the CSM or ISM around the
SN, or a combination of normal dust from CSM and ISM \citep{Foley14}. If we were to use the intrinsic colors 
of the H07 template instead, the $R_V$ obtained would be even lower. Such a low $R_V$ for SN~2010ev has recently 
also been constrained by \citet{Burns14} who applied a detailed Baysian analysis to a large sample of SN~Ia
light curves. They obtained 
$R_V=1.54^{+0.57}_{-0.59}$ and $A_V=0.50^{+0.17}_{-0.19}$ which yields $E(B-V)=0.32$,  consistent with our
approach. One can see that the $u'$ band is crucial to differentiate between different reddening law values.
The NIR could help to constrain these estimates further, however we do not have NIR photometry.\\ 
\indent We did similar fits to data at other epochs, in order 
to investigate the evolution of the reddening law. We do not find any significant change for $R_V$ nor $E(B-V)$
between $-4$ and +15 days from maximum. This argues for no evolution and therefore no nearby dust. We note that 
for SN~2014J, a highly redenned SN, there is an increase of $\sim0.4$ in $R_V$ in the same time range \citep{Foley14}.\\
\indent We note that the intrinsic color of SN~2010ev using the observed colors measured by SiFTO and our host 
reddening estimate is: $(B-V)_{int}=(B-V)-E(B-V)\simeq0.06$ which is slightly redder than the average intrinsic SN~Ia 
color of $(B-V)_{int}\simeq0$. This is consistent with the idea that HVG SNe might be redder both 
intrinsically and because of host extinction \citep{Foley11a}.

\subsection{Narrow absorption features of intervening material}

\begin{figure}[hbtp]
\centering
\includegraphics[width=8.7cm]{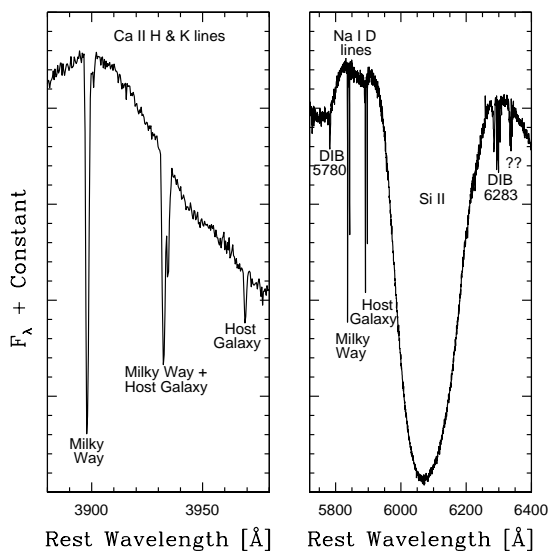}
\caption{Inset of medium-resolution spectra of SN~2010ev showing characteristic narrow absorption 
lines from intervening material in the line of sight.}
\label{dibs}
\end{figure}

The spectra of SN~2010ev exhibit strong interstellar
narrow absorption lines $\textrm{Na\,{\sc i}\,D}$ and \ion{Ca}{ii} H \& K at the redshift of
the host galaxy and the MW, as well as narrow absorption features that correspond to DIBs at the 
redshift of the host galaxy ($\lambda5780$ and $\lambda6283$). Typically, DIBs tend to be seen
in the spectra of stars \citep{Herbig95} or supernovae \citep[e.g.][]{Welty14} reddened by interstellar
dust, giving further evidence for the strong extinction inferred for SN~2010ev in the previous section. 
The identification of these lines was made using the diffuse interstellar band 
catalog\footnote{http://leonid.arc.nasa.gov/DIBcatalog.html} \citep{Jenniskens94}.
Figure~\ref{dibs} shows some of these lines and another unknown narrow line complex in the red part
of $\lambda$6283 \AA\, which we have not identified. \\

\indent Temporal evolution of these lines could signify changes in ionization balance induced by 
the SN radiation field as shown for 
$\textrm{Na\,{\sc i}\,D}$ \citep{Patat07,Blondin09,Simon09,Sternberg13}, for \ion{K}{i} \citep{Graham15} and possibly for DIBs as well
\citep{Milisavljevic14}. We investigate this by analyzing the temporal evolution of the pEW of these
lines. We measure them simply by tracing a straight line along the absorption feature to mimic the 
pseudo-continuum flux, and then estimate the error measuring the pEW many times, changing the trace of the continuum.
Using these multiple measurements we calculate a mean and take the standard deviation to be the error 
on that measurement. The absorption features of $\textrm{Na\,{\sc i}\,D}$ and \ion{Ca}{ii} H \& K from 
the MW and the host galaxy do not seem to evolve with time. Regarding the DIBs, we find no evidence of 
evolution for $\lambda$5780 but a slight decrease for $\lambda$6283.

\indent Recent studies of moderate-resolution absorption lines of $\textrm{Na\,{\sc i}\,D}$ have revealed an 
interesting excess of blueshifted absorptions towards SNe~Ia, 
suggestive of pre-SN outflows
 \citep{Sternberg11,Maguire13,Phillips13}. The SNe that show blueshifted
absorption seem to relate with the strength of 
the $\textrm{Na\,{\sc i}\,D}$ absorption, as well as with the color of the SN \citep{Maguire13}. Having strong 
absorption features and red colors, one could expect SN~2010ev to present these shifts. However, we do not find 
any evidence for blueshift in any of our narrow absorption lines (in agreement with \citealt{Phillips13}) 
arguing for a low amount of CSM in this SN, nevertheless, this may also be due to a lack
of sufficient spectral resolution to confidently rule out such a feature.\\
\indent Besides giving us possible information about the CSM interaction, the strength of the line
can be a useful diagnostic of the amount of absorption by intervening material in the line of sight.
The \ion{Na}{i} D doublet has been used in the past as an indicator of the host galaxy extinction 
\citep{Turatto03,Poznanski12} although its validity has been put into question 
\citep{Poznanski11,Phillips13}. 
Using the method introduced by 
\citet{Phillips13} with column densities of \ion{Na}{i} D, one gets:
$E(B-V)_{\mathrm{MW}}=0.09$ (using $R_V=3.1$) and $E(B-V)_{\mathrm{Host}}=0.25$ (using $R_V=1.54$).\\
\indent For the host galaxy, we can additionally use independent constraints on the reddening from the DIBs.
Using the relations between the EW of the DIBs and $E(B-V)$ proposed by \citet{Luna08} for post-AGB
stars, we estimate a reddening for $\lambda5780$ of $E(B-V)_{\mathrm{Host}}=0.53\pm0.09$ and 
for $\lambda6283$ of $E(B-V)_{\mathrm{Host}}=0.50\pm0.04$. These values are too large compared with
other photometric and spectroscopic estimates, as much as a factor of $\sim2$, as also recently 
shown by \citet{Welty14}. For the host, there is another narrow line that we can use, \ion{K}{i},
with the relation found by \citet{Phillips13}, yielding $E(B-V)_{\mathrm{Host}}=0.16$ (using $R_V=1.54$).
These results are summarized in table~\ref{ebv-table}.
The color excesses obtained with photometric analysis give consistent results, while the narrow absorption features 
agree only for the more recent studies of \citet{Phillips13}.

\section{Discussion}
\label{s4}

\subsection{Bolometric luminosity and Nickel mass}
\label{bolom}

\begin{figure}[hbtp]
\centering
\includegraphics[width=8.7cm]{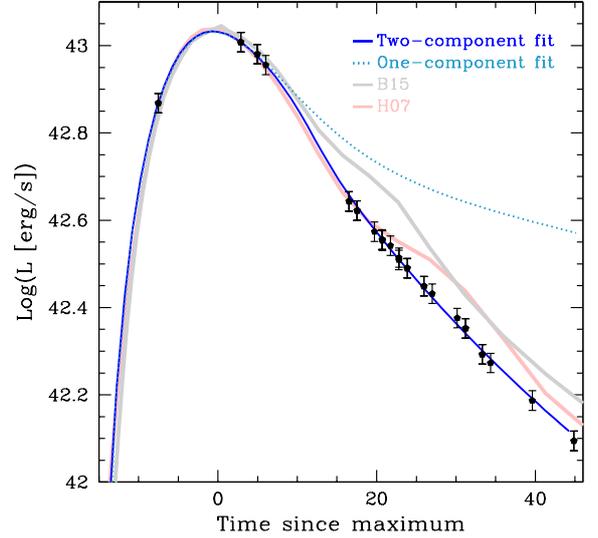}
\caption{Bolometric light-curve of SN~2010ev. 
Circles denote the optical pseudo-bolometric light-curve, the solid blue line is the two-component fit (see text), 
the dotted cyan line is the one-component fit (simple model),
and the red and grey lines are the pseudo-bolometric light-curves for the H07 template and B15 model normalized to the 
peak bolometric luminosity of SN~2010ev.} 
\label{bol-lc}
\end{figure}

In this section we calculate the bolometric light-curve of SN~2010ev, a valuable tool to describe the
general properties of the SN, and to infer characteristics of the explosion and progenitor. We assume the
reddening law of $R_V=1.54$ for the host, as previously calculated, with an extinction of $E(B-V)_{Host}=0.25$. 
We used a distance of $29.88\pm8.11$ Mpc taken from a mean of several methods from NED. Since we have 
enough photometric optical coverage but no NIR nor UV photometry, we can obtain a ``pseudo-bolometric"
light-curve by integrating all $u'Bg'Vr'Ri'I$ photometry corrected for MW and host extinctions. 
The pseudo-bolometric light curve is shown in circles in Figure~\ref{bol-lc}. Errors are calculated 
integrating the photometry taking into account their respective errors. 
We also obtain the bolometric light-curve for the H07 template and the B15 model (solid 
and dashed lines in Figure~\ref{bol-lc}) from the integration of their SEDs in the same wavelength range covered by the filters.\\
\indent The bolometric luminosity at maximum is $L_{\mathrm{bol}}=(1.54\pm0.07)\times10^{43}$erg s$^{-1}$. 
A simple estimate of the $^{56}$Ni mass synthesized during 
the explosion can be obtained using Arnett's rule \citep{Arnett82}.
Adopting a rise-time, i.e. the duration
from explosion to maximum light, of $t_r=16.00\pm0.21$ days given the stretch of the SN \citep{Gonzalez12} 
we obtain $0.56\pm0.10 M_{\sun}$ of $^{56}$Ni.\\
\indent To model the entire bolometric light curve, we employ a modified version of the approach by 
\citet{Arnett82}: the work by \citet{Maeda03} shows that a two-component model is necessary to fit both the 
optically thick and thin regimes.\footnote{The two-component model was applied to hypernovae by \citet{Maeda03},
however it is more generally applicable to any event powered predominantly by the decay of $^{56}$Ni.}
Therefore, in addition to the synthesized $^{56}$Ni mass ($M_\text{Ni}$),
the total ejected mass ($M_\text{ej}$), and the kinetic energy ($E_\text{k}$), two new physical parameters are
invoked: the fraction of mass ($f_\text{M}$) and energy ($f_\text{E}$) of the inner component. Moreover, 
in addition to these two physical components, we define two different regimes in the temporal evolution
of the luminosity: the optically thick phase (around maximum light) and the optically thin 
(or nebular) phase. In our modeling procedure, the transition will lie at the epoch when the optically
thick model does not provide a good fit anymore. Usually it would overestimate the luminosity 30 to 
40 days after the explosion.\\
\indent The optically thick phase was modeled as if only the outer component was contributing to the total 
observed luminosity according to \citet{Valenti08}.
As suggested by \citet{Pignata08}, we measure the velocity at maximum absorption of \ion{S}{ii} $\lambda 5460\AA$ 
and obtain $5270\pm110$ km s$^{-1}$. Using SYNOW \citep{Branch06}, we obtain 
a range of acceptable velocity matches between $\sim5000-10000$ km s$^{-1}$. Thus, we use $7500\pm2500$ km s$^{-1}$
and model the error with a Monte Carlo simulation.\\
\indent The final physical parameters we obtain are $M_{Ni}=0.51\pm0.01 M_\odot$, with 
$t_0 = -15.8 \pm 0.1$ days, and $M_{ej} = 1.2 \pm 0.5 M_\odot$. The other parameters are presented in Table~\ref{bol}

\begin{table}
\caption{Main bolometric parameters of SN~2010ev\label{bol}}
\centering
\begin{tabular}{l l}
\hline
\hline
$M_{Ni}$                      & $0.51\pm0.01 M_\odot$ \\
$t_0$			      & $-15.8\pm 0.1$ \\
$M_{ej}$		      & $1.2\pm 0.5 M_\odot$ \\
$\tau_m \propto M_{ej}^3/E_k$ & $( 1.78^{+0.31}_{-0.16} )\times 10^{-51} M_\odot^3 erg^{-1}$ \\
$f_M$                         & $0.08\pm 0.01$ \\
$f_E$                         & $0.01\pm0.01$ \\
$\upsilon_\text{ph}$          & $7500 \pm 2500$ km s$^{-1}$ \\
$E_k$                         & $( 1.1 \pm 1.2 )\times 10^{51} erg$\\
\end{tabular}
\end{table}

\indent The nickel mass that we obtain is quite standard for a normal stretch $s\sim1$ SN~Ia. The B15 model
has $0.51 M_{\sun}$ of synthetized nickel in remarkable agreement with our estimate. This value is very 
similar to the nickel mass inferred by \citet{Stehle05} for SN~2002bo. \citet{Scalzo14a} find a 
relation between peak $M_B$ and $M_{Ni}$ . Based on their relations, from our observed 
extinction-corrected $M_B$, we get $M_{Ni}\sim0.26 M_{\sun}$, clearly below our calculation, even if we
assume a standard $R_V=3.1$ host reddening law. 
This could be due to the fact that the relations presented in \citet{Scalzo14a} are
for SNe~Ia with normal colors, and are not suited for HVG SNe~Ia. \\
\indent The ejecta mass we obtain is consistent within the errors with a Chandrasekhar explosion, although
a sub-Chandrasekhar event is possible. If this were the case, it would support the idea that a fraction of SNe~Ia 
explodes below the Chandrasekhar mass \citep{Stritzinger06a, Scalzo14b}. Using our stretch and the relation of
\citet{Scalzo14a}, we obtain $M_{ej}\sim1.24 M_{\sun}$.

\subsection{Comparison with other SNe and templates}
 
The analysis done in the previous sections shows that the properties of SN~2010ev 
such as high velocities and their fast gradient, evolution of the ${\cal R}$(\ion{Si}{ii}), 
line widths, and overall match with the spectra at different epochs, are in better agreement 
with the B15 model and HVG SNe. Since the B15 model is a good match to SN~2002bo, 
a similar object to our SN, we expect all of them (SN~2002bo-like objects) to have similar physical
parameters like density profile, explosion energy, nickel mass and nucleosynthetic yields
(see \citealt{Blondin15}).  \\

\indent The differences between these two groups of objects (HVG and LVG), 
best represented by the H07 template and the B15 model, could suggest two different 
explosion mechanisms in normal SNe~Ia. On one side, the Chandrasekhar mass
delayed detonation model can fit the HVG SNe quite well, while on the other hand,
the more common LVG SNe are not well matched and could be explained by alternative models such as
pulsational-delayed detonations \citep{Dessart14}.
The colors at maximum and their evolution
also differ considerably. However, both groups have similar light-curve widths 
and only small differences in $B$-band magnitude. This shows that the diversity of SNe~Ia 
goes beyond a simple parameter related to brightness and light-curve width, 
and a further degree of complexity gives rise to spectral 
and color differences.

\subsection{Extinction and reddening}

It has become clear in recent years that for some SNe~Ia we infer extremely low $R_V$ values
(even lower than 2) in the line of sight to their host galaxies that are atypical of the MW.  
This has been shown for average reddening laws of SN populations \citep{Wang09} and
for individual SNe \citep[e.g.][]{Burns14}.
SN~2010ev is one of these SNe. The reddening is inferred from photometric 
and spectroscopic studies. To do this, one should not assume a global intrinsic color for all
SNe~Ia: here we use the B15 model which is well matched to another similar HVG SN, SN~2002bo. 
The unusual reddening laws for SN~2010ev and other SNe are confirmed through other studies such as \citet{Phillips13} 
and also through independent polarimetric analysis \citep{Zelaya15,Kawabata14}. As a matter of fact, 
\citet{Zelaya15} measure a continuum polarization for SN~2010ev of $(1.67\pm0.65)\%$,
higher than for normal unreddened SNe~Ia, and peaking  at
$\lesssim$3500\AA\ implying an $R_V\lesssim 2$ \citep{Hough87,Wang03a,Patat09}. 

These SNe with low $R_V$ values are redder, they suffer more extinction, and have spectra that present
narrow absorption lines and DIBs, typically attributed
to the same intervening material. This poses the question of whether the deviation from the standard 
MW reddening law is a common feature of SN host environments.
However, a recent study of the reddening of the environment of the redenned SN~2014J \citep{Hutton15} shows 
that the $R_V$ values in the vicinity of the SN do not have the atypical value found with the SN 
data of $R_V\sim1.4$ \citep{Foley14,Amanullah14}, arguing 
for an effect of the SN radiation on nearby material.
Furthermore, the observation of an excess population of SNe~Ia
with blueshifted \ion{Na}{i} D absorption \citep{Sternberg11} has been explained through
outflowing nearby material instead of galactic winds \citep{Park15}, especially since the same 
excess is not observed for CC~SNe.
SNe with low $R_V$ values are redder at maximum due to extinction but 
possibly also intrinsically, and they have characteristic ejecta properties different from the LVG objects. 
These objects are generally HVG SNe, which show a correlation between their 
high velocities decline rates at early epochs and the redshifted nebular velocities at late phases. This
relation between intrinsic and extrinsic properties has encouraged the idea that perhaps the 
material in the line of sight stems from the progenitor itself. Recent searches for light-echoes
from CSM for a large fraction of SNe~Ia in optical spectra \citep{Marino15} or NIR photometry
\citep{Maeda14} defy the CSM light-echo hypothesis. Either another mechanism like CSM dust 
sublimation is at hand or simply low $R_V$'s originate from ISM (perhaps patchy ISM \citealt{Forster13}).
Furthermore, the majority of these SNe (with few exceptions) do not show any variation of narrow absorption 
lines nor of the continuum polarization, which in principle are typical signatures of CSM.  


\section{Conclusions}
\label{s5}

We presented an analysis of SN~2010ev spectra and photometry ranging from $-7.5$ to 289.5 days around the time of 
$B$-band maximum. SN~2010ev has a light-curve width typical of normal SN~Ia, but with
high reddening in the line of sight. The presence of strong
Na I D features invariant in time and DIBs located at $\sim5780$ and $\sim6283$ \AA\ suggest a high extinction possibly
due to ISM of the host galaxy. The E$(B-V)$ is calculated with different methods and our best estimate is 
$E(B-V)_{Host}=0.25\pm0.05$ with an $R_V=1.54\pm0.65$. \\
\indent  In general, the spectral evolution of SN~2010ev shows similarities to the 
high velocity gradient (HVG) group \citep{Benetti05}, with a velocity gradient value of 
$\dot v_{\mathrm{Si}}=164\pm7$ km s$^{-1}$ d$^{-1}$. Within the classification presented by \citet{Branch06},
SN~2010ev is placed in the BL group, which is almost equivalent to HVG. The early-time spectra, 
both in the optical and NIR ranges, show characteristic P-Cygni profiles 
of \ion{Si}{ii}, \ion{Ca}{ii}, 
\ion{S}{ii}, \ion{Mg}{ii}, and \ion{Fe}{ii}.  
We found no evidence of either High Velocity Features (HVFs) or \ion{C}{ii} lines, due to the lack of very early-time spectra.
The nebular spectra show the redshifted \ion{[Fe}{ii]} $\lambda7155$ and
\ion{[Ni}{ii]} $\lambda7378$ features, which may suggest an asymmetry during the explosion \citep{Maeda10}. 
The estimated nebular velocity  $v_{neb}=2145\pm225$ km s$^{-1}$ is consistent with that of HVG SNe. The spectral and photometric evolution is 
also similar to other HVG SNe~Ia such as SN~2002bo and SN~2002dj. \\
\indent  The bolometric analysis gives a nickel mass of $0.51\pm0.01 M_\odot$ in agreement with $0.51M_\odot$ 
from the Chandrasekhar delayed-detonation model of \citet{Blondin15}, and an ejected mass of $1.2\pm 0.5 M_\odot$, 
also consistent with a standard Chandrasekhar mass explosion.
This SN is a further member of the group of SNe~Ia with normal light-curve decline and standard inferred
nickel mass, yet reddened by dust with an unusual low $R_V$. It also has high velocity gradient and redshifted
nebular velocities. There seems to be a coincidence of intrinsic and extrinsic properties for certain SNe~Ia that 
necessitates an explanation and may be key to understanding the SN~Ia progenitor and explosion mechanism.

\begin{acknowledgements}

We thank the annonymous referee for the useful suggestions.
C.P.G, S.G., G.P., M.H., F.B. acknowledge support by projects IC120009 ``Millennium Institute of Astrophysics (MAS)" and P10-064-F
``Millennium Center for Supernova Science'' of the Iniciativa Cient\'ifica Milenio del Ministerio Econom\'ia, Fomento y Turismo 
de Chile.  S.G. aknowledgeds support by CONICYT through FONDECYT grant 3130680. G.P. acknowledges support by the Proyecto 
FONDECYT 11090421.  M.S.  gratefully  acknowledges the generous support provided by the Danish Agency for Science and 
Technology and Innovation  realized through a Sapere Aude Level 2 grant. 
S.T. is supported by the Transregional Collaborative Research Centre TRR 33 ``The Dark Universe'' of the DFG
This research has made use of the NASA/IPAC Extragalactic Database (NED) which is operated by the Jet Propulsion 
Laboratory, California Institute of Technology, under contract with the National Aeronautics.
     
\end{acknowledgements}

\appendix

\section{Photometric tables}
\label{ap1}

\begin{table*}[htb]
\tiny
\centering
\caption{ $BVRI$ and $u'g'r'i'z'$ magnitudes of the local photometric standard stars in the field 
of SN 2010ev (Figure. 1), obtained with PROMPT1, PROMPT3, and PROMPT5. }
\label{stars}
\tabcolsep=0.075cm
\resizebox{\textwidth}{!}{%
\begin{tabular}{@ {}l l l l l l l l l l l l l l l l l l l}
\\
\hline \hline
ID & \multicolumn{1}{c}{B} & \multicolumn{1}{c}{V} & \multicolumn{1}{c}{R} & \multicolumn{1}{c}{I} 
& \multicolumn{1}{c}{$u'$} & \multicolumn{1}{c}{$g'$} & \multicolumn{1}{c}{$r'$} & \multicolumn{1}{c}{$i'$} & \multicolumn{1}{c}{$z'$} \\
\hline            
 1  & $15.74\pm0.02$ & $15.10\pm0.03$ & $14.75\pm0.02$ & $14.38\pm0.02$ & 
    $16.70\pm0.07$ & $15.41\pm0.01$ & $14.97\pm0.01$ & $14.81\pm0.01$ & $14.79\pm0.01$ \\
 2  & $17.08\pm0.04$ & $16.20\pm0.04$ & $15.71\pm0.03$ & $15.26\pm0.02$ & 
    $18.50\pm0.09$ & $16.62\pm0.01$ & $15.94\pm0.01$ & $15.70\pm0.01$ & $15.62\pm0.01$ \\   
 3  & $14.39\pm0.01$ & $13.77\pm0.03$ & $13.43\pm0.02$ & $13.07\pm0.02$ & 
    $15.36\pm0.03$ & $14.06\pm0.01$ & $13.65\pm0.01$ & $13.51\pm0.01$ & $13.49\pm0.01$ \\ 
 4  & $16.83\pm0.04$ & $15.74\pm0.034$ & $15.19\pm0.03$ & $14.63\pm0.03$ & 
    $18.76\pm0.01$ & $16.27\pm0.01$ & $15.45\pm0.01$ & $15.12\pm0.01$ & $14.97\pm0.01$ \\  
 5  & $16.90\pm0.03$ & $15.93\pm0.03$ & $15.43\pm0.03$ & $14.91\pm0.02$ & 
                     & $16.41\pm0.01$ & $15.68\pm0.01$ & $15.38\pm0.01$ & $15.26\pm0.01$ \\  
 6  & $17.42\pm0.04$ & $16.42\pm0.03$ & $15.84\pm0.03$ & $15.28\pm0.02$ & 
                     & $16.94\pm0.02$ & $16.09\pm0.017$ & $15.77\pm0.01$ & $15.61\pm0.01$ \\  
 7  &                  & $17.05\pm0.04$ & $16.58\pm0.02$ & $16.11\pm0.04$ & 
    $19.30\pm0.01$ & $17.50\pm0.01$ & $16.82\pm0.01$ & $16.57\pm0.01$ & $16.46\pm0.01$ \\  
 8  & $16.14\pm0.02$ & $15.35\pm0.03$ & $14.93\pm0.02$ & $14.51\pm0.02$ & 
    $17.33\pm0.03$ & $15.75\pm0.01$ & $15.18\pm0.01$ & $14.97\pm0.01$ & $14.89\pm0.01$ \\  
 9  & $18.10\pm0.08$ & $17.45\pm0.04$ & $17.10\pm0.04$ & $16.72\pm0.02$ & 
                     & $17.80\pm0.01$ & $17.33\pm0.01$ & $17.15\pm0.01$ & $17.11\pm0.01$ \\  
10  & $17.53\pm0.03$ & $16.83\pm0.02$ & $16.46\pm0.03$ & $16.06\pm0.03$ & 
                     & $17.21\pm0.02$ & $16.68\pm0.01$ & $16.49\pm0.01$ & $16.43\pm0.02$ \\  
11  & $16.08\pm0.01$ & $15.44\pm0.04$ & $15.09\pm0.02$ & $14.73\pm0.02$ & 
    $16.98\pm0.00$ & $15.76\pm0.02$ & $15.30\pm0.01$ & $15.15\pm0.01$ & $15.12\pm0.01$ \\  
12  & $17.47\pm0.01$ & $16.60\pm0.03$ & $16.15\pm0.03$ & $15.70\pm0.03$ & 
    $18.86\pm0.06$ & $17.04\pm0.01$ & $16.37\pm0.01$ & $16.15\pm0.01$ & $16.07\pm0.03$ \\  
13  &                  &                  &                  &                  & 
                     & $17.55\pm0.00$ & $16.92\pm0.01$ & $16.69\pm0.01$ & $16.62\pm0.04$ \\  
14  & $13.78\pm0.02$ & $13.21\pm0.03$ & $12.88\pm0.02$ & $12.56\pm0.02$ & 
    $14.70\pm0.03$ & $13.49\pm0.01$ & $13.12\pm0.01$ & $13.00\pm0.01$ & $13.00\pm0.01$ \\  
15  & $15.42\pm0.02$ & $14.81\pm0.03$ & $14.46\pm0.02$ & $14.12\pm0.02$ & 
    $16.46\pm0.02$ & $15.13\pm0.01$ & $14.69\pm0.01$ & $14.55\pm0.01$ & $14.53\pm0.01$ \\  
16  & $16.27\pm0.02$ & $15.41\pm0.03$ & $14.94\pm0.02$ & $14.47\pm0.02$ & 
    $17.47\pm0.01$ & $15.85\pm0.01$ & $15.20\pm0.01$ & $14.94\pm0.01$ & $14.84\pm0.02$ \\  
17  & $14.38\pm0.03$ &                  & $12.28\pm0.02$ & $11.59\pm0.02$ & 
                     & $13.65\pm0.01$ & $12.56\pm0.00$ & $12.10\pm0.01$ &		     \\ 
18  & $14.77\pm0.02$ & $14.11\pm0.03$ & $13.73\pm0.02$ & $13.32\pm0.02$ &
    $15.63\pm0.02$ &                  & $13.95\pm0.00$ & $13.76\pm0.00$ & 		     \\
19  & $16.93\pm0.02$ &                  & $15.24\pm0.03$ & $14.69\pm0.02$ & 
    $18.77\pm0.06$ & $16.38\pm0.01$ & $15.52\pm0.01$ &                  & $15.03\pm0.01$ \\  
20  & $16.36\pm0.03$ &                  & $15.20\pm0.03$ & $14.79\pm0.02$ & 
    $17.42\pm0.07$ & $15.96\pm0.02$ & $15.43\pm0.00$ & $15.23\pm0.00$ & 		     \\
21  & $12.26\pm0.02$ & $11.77\pm0.03$ & $11.48\pm0.02$ &                  & 
    $13.16\pm0.03$ &                  & $11.67\pm0.01$ & $11.60\pm0.00$ & 		     \\ 
\hline \hline
\end{tabular}}
\end{table*}

\begin{landscape}
\begin{table}
\centering
\tiny
\caption{$BVRI$ and Sloan photometry of SN 2010ev with PROMPT1, PROMPT3 and PROMPT5 telescopes.}
\label{photom}
\tabcolsep=0.075cm
\begin{tabular}{@ {}l l l l l l l l l l l l}
\\
\hline \hline
UT date & M.J.D. & Phase$^{\star}$ & \multicolumn{1}{c}{B} & \multicolumn{1}{c}{V} & \multicolumn{1}{c}{R} & \multicolumn{1}{c}{I} &
 \multicolumn{1}{c}{$u'$} & \multicolumn{1}{c}{$g'$} & \multicolumn{1}{c}{$r'$} & \multicolumn{1}{c}{$i'$} & \multicolumn{1}{c}{$z'$}\\
\hline
2010/06/30  &  55377.08 & -7.5  & $15.45\pm0.01$ & $15.21\pm0.01$ & $15.01\pm0.01$ & $15.02\pm0.01$ 
				& $16.53\pm0.06$ & $15.34\pm0.01$ & $15.12\pm0.01$ & $15.50\pm0.01$ & $15.42\pm0.02$ \\
2010/07/10  &  55386.98 & 2.4   & $15.01\pm0.01$ & $14.64\pm0.01$ & $14.48\pm0.01$ & $14.68\pm0.01$ 
				& $16.19\pm0.05$ & $14.80\pm0.01$ & $14.62\pm0.01$ & $15.24\pm0.01$ & $15.16\pm0.01$ \\
2010/07/12  &  55388.60 & 4.0   &                & $14.69\pm0.01$ & $14.50\pm0.01$ & $14.80\pm0.01$ 
				& $16.39\pm0.05$ & $14.88\pm0.01$ & $14.64\pm0.01$ & $15.27\pm0.01$ & $15.23\pm0.01$ \\
2010/07/12  &  55388.96 & 4.4   & $15.08\pm0.01$ & $14.65\pm0.01$ & $14.50\pm0.01$ & $14.76\pm0.01$ \\
2010/07/13  &  55389.98 & 5.4   &                & $14.71\pm0.01$ & $14.57\pm0.01$ & $14.85\pm0.01$ \\
2010/07/23  &  55399.99 & 15.4  & $16.19\pm0.02$ & $15.3\pm0.01$  & $15.14\pm0.01$ & $15.15\pm0.01$ 
				& $17.79\pm0.09$ & $15.71\pm0.01$ &                &                &                   \\
2010/07/24  &  55400.95 & 16.3  &                & $15.27\pm0.04$ & $15.14\pm0.03$ & $15.12\pm0.04$ 
				&                &                & $15.26\pm0.02$ & $15.88\pm0.05$ & $15.43\pm0.04$ \\
2010/07/26  &  55403.03 & 18.4  & $16.61\pm0.03$ & $15.38\pm0.01$ & $15.14\pm0.01$ & $15.03\pm0.02$ \\
2010/07/27  &  55403.97 & 19.4  & $16.70\pm0.02$ & $15.39\pm0.03$ & $15.13\pm0.02$ & $15.03\pm0.02$ 
				& $18.40\pm0.12$ & $15.96\pm0.02$ & $15.28\pm0.01$ & $15.75\pm0.02$ & $15.54\pm0.03$ \\
2010/07/27  &  55403.97 & 19.4  & $16.70\pm0.02$ & $15.44\pm0.01$ & $15.15\pm0.01$ & $15.04\pm0.01$ \\
2010/07/28  &  55404.96 & 20.4  & $16.77\pm0.06$ & $15.48\pm0.02$ & $15.15\pm0.01$ & $15.01\pm0.01$ \\
2010/07/28  &  55404.96 & 20.4  & $16.79\pm0.03$ & $15.46\pm0.02$ &                &                  
				&                &                &                &              & $15.59\pm0.03$ \\
2010/07/29  &  55405.98 & 21.4  & $16.95\pm0.01$ & $15.58\pm0.01$ & $15.21\pm0.01$ & $15.02\pm0.01$ 
				& $18.47\pm0.10$ & $16.16\pm0.01$ & $15.30\pm0.02$ & $15.67\pm0.03$ &                  \\
2010/07/29  &  55405.98 & 21.4  & $16.91\pm0.03$ & $15.55\pm0.01$ &                &                  \\
2010/07/30  &  55406.98 & 22.4  & $17.06\pm0.02$ & $15.63\pm0.01$ & $15.21\pm0.01$ & $15.00\pm0.01$
				& $18.66\pm0.09$ & $16.22\pm0.02$ &                &                  &                  \\
2010/07/30  &  55406.98 & 22.4  & $17.08\pm0.02$ & $15.60\pm0.01$ &                &                  \\
2010/08/01  &  55408.99 & 24.4  & $17.29\pm0.02$ & $15.76\pm0.01$ & $15.26\pm0.01$ & $14.97\pm0.01$
				&                & $16.40\pm0.01$ & $15.39\pm0.02$ & $15.69\pm0.03$ & $15.29\pm0.01$ \\
2010/08/02  &  55409.97 & 25.4  &                & $15.80\pm0.01$ & $15.30\pm0.01$ & $14.98\pm0.01$ \\
2010/08/05  &  55412.98 & 28.4  &                & $15.94\pm0.03$ &                &                  \\
2010/08/06  &  55413.98 & 29.4  & $17.64\pm0.04$ &                & $15.52\pm0.01$ & $15.07\pm0.01$ 
				&                & $16.75\pm0.01$ & $15.67\pm0.01$ & $15.74\pm0.01$ & $15.35\pm0.01$ \\
2010/08/08  &  55415.98 & 31.4  & $17.79\pm0.04$ & $16.24\pm0.01$ & $15.64\pm0.01$ & $15.24\pm0.01$ 
				&                & $16.88\pm0.01$ & $15.84\pm0.01$ & $15.88\pm0.02$ &                  \\
2010/08/09  &  55416.97 & 32.4  &	         & 	   	  &		   &			
				&                &                &                & $16.02\pm0.05$ &                  \\ 				 
2010/08/14  &  55421.98 & 37.4  &                &                & $16.01\pm0.01$ &                  \\
2010/08/19  &  55426.98 & 42.4  & $18.30\pm0.06$ & $16.70\pm0.02$ & $16.21\pm0.01$ & $16.05\pm0.06$ 
				&                & $17.35\pm0.02$ & $16.37\pm0.01$ & $16.56\pm0.04$ &                  \\
2010/12/29  &  55560.03 & 175.4 &                & $19.84\pm0.02$ & $20.16\pm0.01$ & $20.11\pm0.03$ 
            &                   &                &                &                &                  \\
2011/03/28  &  55649.15 & 264.6 & $20.85\pm0.04$ & $20.70\pm0.44$ &                &
                                &                & $20.49\pm0.21$ & $21.35\pm0.21$ &                &              \\
2011/04/03  &  55655.03 & 270.4 & $20.99\pm0.04$ & $21.20\pm0.03$ & $21.55\pm0.05$ & $21.12\pm0.04$  
            &                   &                &                &                &                  \\   
2011/04/23  &  55674.08 & 289.5 & $21.27\pm0.23$ & $21.17\pm0.10$ & $21.33\pm0.12$ & $20.93\pm0.13$		       	
	    &                   & $20.88\pm0.06$ & $21.86\pm0.18$ & $21.08\pm0.09$ &                  \\
\hline \hline
\end{tabular}
\begin{list}{}{}
\item[$^{\star}$] Relative to B$_{max}$ (MJD$=2455384.60)$
\end{list}
\end{table}
\end{landscape}


\begin{thebibliography}{111}
\expandafter\ifx\csname natexlab\endcsname\relax\def\natexlab#1{#1}\fi

\bibitem[{{Altavilla} {et~al.}(2007){Altavilla}, {Stehle}, {Ruiz-Lapuente},
  {Mazzali}, {Pignata}, {Balastegui}, {Benetti}, {Blanc}, {Canal},
  {Elias-Rosa}, {Goobar}, {Harutyunyan}, {Pastorello}, {Patat}, {Rich},
  {Salvo}, {Schmidt}, {Stanishev}, {Taubenberger}, {Turatto}, \&
  {Hillebrandt}}]{Altavilla07}
{Altavilla}, G., {Stehle}, M., {Ruiz-Lapuente}, P., {et~al.} 2007, \aap, 475,
  585

\bibitem[{{Amanullah} \& {Goobar}(2011)}]{Amanullah11}
{Amanullah}, R. \& {Goobar}, A. 2011, \apj, 735, 20

\bibitem[{{Amanullah} {et~al.}(2014)}]{Amanullah14}
{Amanullah}, R. {et~al.} 2014, \apjl, 788, L21

\bibitem[{{Arnett}(1982)}]{Arnett82}
{Arnett}, W.~D. 1982, \apj, 253, 785

\bibitem[{{Benetti} {et~al.}(2005){Benetti}, {Cappellaro}, {Mazzali},
  {Turatto}, {Altavilla}, {Bufano}, {Elias-Rosa}, {Kotak}, {Pignata}, {Salvo},
  \& {Stanishev}}]{Benetti05}
{Benetti}, S., {Cappellaro}, E., {Mazzali}, P.~A., {et~al.} 2005, \apj, 623,
  1011

\bibitem[{{Benetti} {et~al.}(2004){Benetti}, {Meikle}, {Stehle}, {Altavilla},
  {Desidera}, {Folatelli}, {Goobar}, {Mattila}, {Mendez}, {Navasardyan},
  {Pastorello}, {Patat}, {Riello}, {Ruiz-Lapuente}, {Tsvetkov}, {Turatto},
  {Mazzali}, \& {Hillebrandt}}]{Benetti04}
{Benetti}, S., {Meikle}, P., {Stehle}, M., {et~al.} 2004, \mnras, 348, 261

\bibitem[{{Blondin} {et~al.}(2015){Blondin}, {Dessart}, \&
  {Hillier}}]{Blondin15}
{Blondin}, S., {Dessart}, L., \& {Hillier}, D.~J. 2015, \mnras, 448, 2766

\bibitem[{{Blondin} {et~al.}(2012){Blondin}, {Matheson}, {Kirshner}, {Mandel},
  {Berlind}, {Calkins}, {Challis}, {Garnavich}, {Jha}, {Modjaz}, {Riess}, \&
  {Schmidt}}]{Blondin12}
{Blondin}, S., {Matheson}, T., {Kirshner}, R.~P., {et~al.} 2012, \aj, 143, 126

\bibitem[{{Blondin} {et~al.}(2009){Blondin}, {Prieto}, {Patat}, {Challis},
  {Hicken}, {Kirshner}, {Matheson}, \& {Modjaz}}]{Blondin09}
{Blondin}, S., {Prieto}, J.~L., {Patat}, F., {et~al.} 2009, \apj, 693, 207

\bibitem[{{Branch} {et~al.}(2009){Branch}, {Dang}, \& {Baron}}]{Branch09}
{Branch}, D., {Dang}, L.~C., \& {Baron}, E. 2009, \pasp, 121, 238

\bibitem[{{Branch} {et~al.}(2006){Branch}, {Dang}, {Hall}, {Ketchum},
  {Melakayil}, {Parrent}, {Troxel}, {Casebeer}, {Jeffery}, \&
  {Baron}}]{Branch06}
{Branch}, D., {Dang}, L.~C., {Hall}, N., {et~al.} 2006, \pasp, 118, 560

\bibitem[{{Burns} {et~al.}(2014){Burns}, {Stritzinger}, {Phillips}, {Hsiao},
  {Contreras}, {Persson}, {Folatelli}, {Boldt}, {Campillay}, {Castell{\'o}n},
  {Freedman}, {Madore}, {Morrell}, {Salgado}, \& {Suntzeff}}]{Burns14}
{Burns}, C.~R., {Stritzinger}, M., {Phillips}, M.~M., {et~al.} 2014, \apj, 789,
  32

\bibitem[{{Burns} {et~al.}(2011){Burns}, {Stritzinger}, {Phillips}, {Kattner},
  {Persson}, {Madore}, {Freedman}, {Boldt}, {Campillay}, {Contreras},
  {Folatelli}, {Gonzalez}, {Krzeminski}, {Morrell}, {Salgado}, \&
  {Suntzeff}}]{Burns11}
{Burns}, C.~R., {Stritzinger}, M., {Phillips}, M.~M., {et~al.} 2011, \aj, 141,
  19

\bibitem[{{Cardelli} {et~al.}(1989){Cardelli}, {Clayton}, \&
  {Mathis}}]{Cardelli89}
{Cardelli}, J.~A., {Clayton}, G.~C., \& {Mathis}, J.~S. 1989, \apj, 345, 245

\bibitem[{{Conley} {et~al.}(2007){Conley}, {Carlberg}, {Guy}, {Howell}, {Jha},
  {Riess}, \& {Sullivan}}]{Conley07}
{Conley}, A., {Carlberg}, R.~G., {Guy}, J., {et~al.} 2007, \apjl, 664, L13

\bibitem[{{Conley} {et~al.}(2008){Conley}, {Sullivan}, {Hsiao}, {Guy},
  {Astier}, {Balam}, {Balland}, {Basa}, {Carlberg}, {Fouchez}, {Hardin},
  {Howell}, {Hook}, {Pain}, {Perrett}, {Pritchet}, \& {Regnault}}]{Conley08}
{Conley}, A., {Sullivan}, M., {Hsiao}, E.~Y., {et~al.} 2008, \apj, 681, 482

\bibitem[{{Dessart} {et~al.}(2014){Dessart}, {Blondin}, {Hillier}, \&
  {Khokhlov}}]{Dessart14}
{Dessart}, L., {Blondin}, S., {Hillier}, D.~J., \& {Khokhlov}, A. 2014, \mnras,
  441, 532

\bibitem[{{Elias-Rosa} {et~al.}(2006)}]{Elias-Rosa06}
{Elias-Rosa}, N. {et~al.} 2006, \mnras, 369, 1880

\bibitem[{{Fitzpatrick}(1999)}]{Fitzpatrick99}
{Fitzpatrick}, E.~L. 1999, \pasp, 111, 63

\bibitem[{{Fitzpatrick} \& {Massa}(2007)}]{Fitzpatrick07}
{Fitzpatrick}, E.~L. \& {Massa}, D. 2007, \apj, 663, 320

\bibitem[{{Folatelli} {et~al.}(2013){Folatelli}, {Morrell}, {Phillips},
  {Hsiao}, {Campillay}, {Contreras}, {Castell{\'o}n}, {Hamuy}, {Krzeminski},
  {Roth}, {Stritzinger}, {Burns}, {Freedman}, {Madore}, {Murphy}, {Persson},
  {Prieto}, {Suntzeff}, {Krisciunas}, {Anderson}, {F{\"o}rster}, {Maza},
  {Pignata}, {Rojas}, {Boldt}, {Salgado}, {Wyatt}, {Olivares E.}, {Gal-Yam}, \&
  {Sako}}]{Folatelli13}
{Folatelli}, G., {Morrell}, N., {Phillips}, M.~M., {et~al.} 2013, \apj, 773, 53

\bibitem[{{Folatelli} {et~al.}(2010){Folatelli}, {Phillips}, {Burns},
  {Contreras}, {Hamuy}, {Freedman}, {Persson}, {Stritzinger}, {Suntzeff},
  {Krisciunas}, {Boldt}, {Gonz{\'a}lez}, {Krzeminski}, {Morrell}, {Roth},
  {Salgado}, {Madore}, {Murphy}, {Wyatt}, {Li}, {Filippenko}, \&
  {Miller}}]{Folatelli10}
{Folatelli}, G., {Phillips}, M.~M., {Burns}, C.~R., {et~al.} 2010, \aj, 139,
  120

\bibitem[{{Foley} {et~al.}(2011){Foley}, {Sanders}, \& {Kirshner}}]{Foley11a}
{Foley}, R.~J., {Sanders}, N.~E., \& {Kirshner}, R.~P. 2011, \apj, 742, 89

\bibitem[{{Foley} {et~al.}(2014)}]{Foley14}
{Foley}, R.~J. {et~al.} 2014, ArXiv e-prints

\bibitem[{{F{\"o}rster} {et~al.}(2012){F{\"o}rster}, {Gonz{\'a}lez-Gait{\'a}n},
  {Anderson}, {Marchi}, {Guti{\'e}rrez}, {Hamuy}, {Pignata}, \&
  {Cartier}}]{Forster12}
{F{\"o}rster}, F., {Gonz{\'a}lez-Gait{\'a}n}, S., {Anderson}, J., {et~al.}
  2012, \apjl, 754, L21

\bibitem[{{F{\"o}rster} {et~al.}(2013){F{\"o}rster}, {Gonz{\'a}lez-Gait{\'a}n},
  {Folatelli}, \& {Morrell}}]{Forster13}
{F{\"o}rster}, F., {Gonz{\'a}lez-Gait{\'a}n}, S., {Folatelli}, G., \&
  {Morrell}, N. 2013, \apj, 772, 19

\bibitem[{{Gall} {et~al.}(2012){Gall}, {Taubenberger}, {Kromer}, {Sim},
  {Benetti}, {Blanc}, {Elias-Rosa}, \& {Hillebrandt}}]{Gall12}
{Gall}, E.~E.~E., {Taubenberger}, S., {Kromer}, M., {et~al.} 2012, \mnras, 427,
  994

\bibitem[{{Gonz{\'a}lez-Gait{\'a}n} {et~al.}(2012){Gonz{\'a}lez-Gait{\'a}n},
  {Conley}, {Bianco}, {Howell}, {Sullivan}, {Perrett}, {Carlberg}, {Astier},
  {et~al.}}]{Gonzalez12}
{Gonz{\'a}lez-Gait{\'a}n}, S., {Conley}, A., {Bianco}, F.~B., {et~al.} 2012,
  \apj, 745, 44

\bibitem[{{Gonz{\'a}lez-Gait{\'a}n} {et~al.}(2014){Gonz{\'a}lez-Gait{\'a}n},
  {Hsiao}, {Pignata}, {F{\"o}rster}, {Guti{\'e}rrez}, {Bufano}, {Galbany},
  {Folatelli}, {Phillips}, {Hamuy}, {Anderson}, \& {de Jaeger}}]{Gonzalez14}
{Gonz{\'a}lez-Gait{\'a}n}, S., {Hsiao}, E.~Y., {Pignata}, G., {et~al.} 2014,
  \apj, 795, 142

\bibitem[{{Goobar}(2008)}]{Goobar08}
{Goobar}, A. 2008, \apjl, 686, L103

\bibitem[{{Goobar} {et~al.}(2014){Goobar}, {Johansson}, {Amanullah}, {Cao},
  {Perley}, {Kasliwal}, {Ferretti}, {Nugent}, {Harris}, {Gal-Yam}, {Ofek},
  {Tendulkar}, {Dennefeld}, {Valenti}, {Arcavi}, {Banerjee}, {Venkataraman},
  {Joshi}, {Ashok}, {Cenko}, {Diaz}, {Fremling}, {Horesh}, {Howell},
  {Kulkarni}, {Papadogiannakis}, {Petrushevska}, {Sand}, {Sollerman},
  {Stanishev}, {Bloom}, {Surace}, {Dupuy}, \& {Liu}}]{Goobar14}
{Goobar}, A., {Johansson}, J., {Amanullah}, R., {et~al.} 2014, ArXiv e-prints

\bibitem[{{Graham} {et~al.}(2015)}]{Graham15}
{Graham}, M.~L. {et~al.} 2015, \apj, 801, 136

\bibitem[{{Guy} {et~al.}(2007)}]{Guy07}
{Guy}, J. {et~al.} 2007, \aap, 466, 11

\bibitem[{{Hamuy} {et~al.}(1996){Hamuy}, {Phillips}, {Suntzeff}, {Schommer},
  {Maza}, {Smith}, {Lira}, \& {Aviles}}]{Hamuy96}
{Hamuy}, M., {Phillips}, M.~M., {Suntzeff}, N.~B., {et~al.} 1996, \aj, 112,
  2438

\bibitem[{{Herbig}(1995)}]{Herbig95}
{Herbig}, G.~H. 1995, \araa, 33, 19

\bibitem[{{Hough} {et~al.}(1987){Hough}, {Bailey}, {Rouse}, \&
  {Whittet}}]{Hough87}
{Hough}, J.~H., {Bailey}, J.~A., {Rouse}, M.~F., \& {Whittet}, D.~C.~B. 1987,
  \mnras, 227, 1P

\bibitem[{{Hoyle} \& {Fowler}(1960)}]{Hoyle60}
{Hoyle}, F. \& {Fowler}, W.~A. 1960, \apj, 132, 565

\bibitem[{{Hsiao} {et~al.}(2007){Hsiao}, {Conley}, {Howell}, {Sullivan},
  {Pritchet}, {Carlberg}, {Nugent}, \& {Phillips}}]{Hsiao07}
{Hsiao}, E.~Y., {Conley}, A., {Howell}, D.~A., {et~al.} 2007, \apj, 663, 1187

\bibitem[{{Hsiao} {et~al.}(2013){Hsiao}, {Marion}, {Phillips}, {Burns},
  {Winge}, {Morrell}, {Contreras}, {Freedman}, {Kromer}, {Gall}, {Gerardy},
  {H{\"o}flich}, {Im}, {Jeon}, {Kirshner}, {Nugent}, {Persson}, {Pignata},
  {Roth}, {Stanishev}, {Stritzinger}, \& {Suntzeff}}]{Hsiao13}
{Hsiao}, E.~Y., {Marion}, G.~H., {Phillips}, M.~M., {et~al.} 2013, \apj, 766,
  72

\bibitem[{{Hutton} {et~al.}(2015){Hutton}, {Ferreras}, \& {Yershov}}]{Hutton15}
{Hutton}, S., {Ferreras}, I., \& {Yershov}, V. 2015, \mnras, 452, 1412

\bibitem[{{Iben} \& {Tutukov}(1984)}]{Iben84}
{Iben}, Jr., I. \& {Tutukov}, A.~V. 1984, \apjs, 54, 335

\bibitem[{{Jenniskens} \& {Desert}(1994)}]{Jenniskens94}
{Jenniskens}, P. \& {Desert}, F.-X. 1994, \aaps, 106, 39

\bibitem[{{Kawabata} {et~al.}(2014){Kawabata}, {Akitaya}, {Yamanaka}, {Itoh},
  {Maeda}, {Moritani}, {Ui}, {Kawabata}, {Mori}, {Nogami}, {Nomoto}, {Suzuki},
  {Takaki}, {Tanaka}, {Ueno}, {Chiyonobu}, {Harao}, {Matsui}, {Miyamoto},
  {Nagae}, {Nakashima}, {Nakaya}, {Ohashi}, {Ohsugi}, {Komatsu}, {Sakimoto},
  {Sasada}, {Sato}, {Tanaka}, {Urano}, {Yamashita}, {Yoshida}, {Arai},
  {Ebisuda}, {Fukazawa}, {Fukui}, {Hashimoto}, {Honda}, {Izumiura}, {Kanda},
  {Kawaguchi}, {Kawai}, {Kuroda}, {Masumoto}, {Matsumoto}, {Nakaoka}, {Takata},
  {Uemura}, \& {Yanagisawa}}]{Kawabata14}
{Kawabata}, K.~S., {Akitaya}, H., {Yamanaka}, M., {et~al.} 2014, \apjl, 795, L4

\bibitem[{{Krisciunas} {et~al.}(2007)}]{Krisciunas07}
{Krisciunas}, K. {et~al.} 2007, \aj, 133, 58

\bibitem[{{Kromer} {et~al.}(2010){Kromer}, {Sim}, {Fink}, {R{\"o}pke},
  {Seitenzahl}, \& {Hillebrandt}}]{Kromer10}
{Kromer}, M., {Sim}, S.~A., {Fink}, M., {et~al.} 2010, \apj, 719, 1067

\bibitem[{{Lair} {et~al.}(2006){Lair}, {Leising}, {Milne}, \&
  {Williams}}]{Lair06}
{Lair}, J.~C., {Leising}, M.~D., {Milne}, P.~A., \& {Williams}, G.~G. 2006,
  \aj, 132, 2024

\bibitem[{{Landolt}(1992)}]{Landolt92}
{Landolt}, A.~U. 1992, \aj, 104, 372

\bibitem[{{Landolt}(2007)}]{Landolt07}
{Landolt}, A.~U. 2007, \aj, 133, 2502

\bibitem[{{Leloudas} {et~al.}(2009){Leloudas}, {Stritzinger}, {Sollerman},
  {Burns}, {Kozma}, {Krisciunas}, {Maund}, {Milne}, {Filippenko}, {Fransson},
  {Ganeshalingam}, {Hamuy}, {Li}, {Phillips}, {Schmidt}, {Skottfelt},
  {Taubenberger}, {Boldt}, {Fynbo}, {Gonzalez}, {Salvo}, \&
  {Thomas-Osip}}]{Leloudas09}
{Leloudas}, G., {Stritzinger}, M.~D., {Sollerman}, J., {et~al.} 2009, \aap,
  505, 265

\bibitem[{{Luna} {et~al.}(2008){Luna}, {Cox}, {Satorre}, {Garc{\'{\i}}a
  Hern{\'a}ndez}, {Su{\'a}rez}, \& {Garc{\'{\i}}a Lario}}]{Luna08}
{Luna}, R., {Cox}, N.~L.~J., {Satorre}, M.~A., {et~al.} 2008, \aap, 480, 133

\bibitem[{{Maeda} {et~al.}(2010{\natexlab{a}}){Maeda}, {Benetti},
  {Stritzinger}, {R{\"o}pke}, {Folatelli}, {Sollerman}, {Taubenberger},
  {Nomoto}, {Leloudas}, {Hamuy}, {Tanaka}, {Mazzali}, \&
  {Elias-Rosa}}]{Maeda10}
{Maeda}, K., {Benetti}, S., {Stritzinger}, M., {et~al.} 2010{\natexlab{a}},
  \nat, 466, 82

\bibitem[{{Maeda} {et~al.}(2003){Maeda}, {Mazzali}, {Deng}, {Nomoto}, {Yoshii},
  {Tomita}, \& {Kobayashi}}]{Maeda03}
{Maeda}, K., {Mazzali}, P.~A., {Deng}, J., {et~al.} 2003, \apj, 593, 931

\bibitem[{{Maeda} {et~al.}(2014){Maeda}, {Nozawa}, {Nagao}, \&
  {Motohara}}]{Maeda14}
{Maeda}, K., {Nozawa}, T., {Nagao}, T., \& {Motohara}, K. 2014, ArXiv e-prints

\bibitem[{{Maeda} {et~al.}(2010{\natexlab{b}}){Maeda}, {Taubenberger},
  {Sollerman}, {Mazzali}, {Leloudas}, {Nomoto}, \& {Motohara}}]{Maeda10a}
{Maeda}, K., {Taubenberger}, S., {Sollerman}, J., {et~al.} 2010{\natexlab{b}},
  \apj, 708, 1703

\bibitem[{{Maeda} {et~al.}(2011)}]{Maeda11}
{Maeda}, K. {et~al.} 2011, \mnras, 413, 3075

\bibitem[{{Maguire} {et~al.}(2013){Maguire}, {Sullivan}, {Patat}, {Gal-Yam},
  {Hook}, {Dhawan}, {Howell}, {Mazzali}, {Nugent}, {Pan}, {Podsiadlowski},
  {Simon}, {Sternberg}, {Valenti}, {Baltay}, {Bersier}, {Blagorodnova}, {Chen},
  {Ellman}, {Feindt}, {F{\"o}rster}, {Fraser}, {Gonz{\'a}lez-Gait{\'a}n},
  {Graham}, {Guti{\'e}rrez}, {Hachinger}, {Hadjiyska}, {Inserra}, {Knapic},
  {Laher}, {Leloudas}, {Margheim}, {McKinnon}, {Molinaro}, {Morrell}, {Ofek},
  {Rabinowitz}, {Rest}, {Sand}, {Smareglia}, {Smartt}, {Taddia}, {Walker},
  {Walton}, \& {Young}}]{Maguire13}
{Maguire}, K., {Sullivan}, M., {Patat}, F., {et~al.} 2013, \mnras, 436, 222

\bibitem[{{Mandel} {et~al.}(2011){Mandel}, {Narayan}, \& {Kirshner}}]{Mandel11}
{Mandel}, K.~S., {Narayan}, G., \& {Kirshner}, R.~P. 2011, \apj, 731, 120

\bibitem[{{Marino} {et~al.}(2015){Marino}, {Gonz{\'a}lez-Gait{\'a}n},
  {F{\"o}rster}, {Folatelli}, {Hamuy}, \& {Hsiao}}]{Marino15}
{Marino}, S., {Gonz{\'a}lez-Gait{\'a}n}, S., {F{\"o}rster}, F., {et~al.} 2015,
  \apj, 806, 134

\bibitem[{{Marion} {et~al.}(2009){Marion}, {H{\"o}flich}, {Gerardy}, {Vacca},
  {Wheeler}, \& {Robinson}}]{Marion09}
{Marion}, G.~H., {H{\"o}flich}, P., {Gerardy}, C.~L., {et~al.} 2009, \aj, 138,
  727

\bibitem[{{Marion} {et~al.}(2003){Marion}, {H{\"o}flich}, {Vacca}, \&
  {Wheeler}}]{Marion03}
{Marion}, G.~H., {H{\"o}flich}, P., {Vacca}, W.~D., \& {Wheeler}, J.~C. 2003,
  \apj, 591, 316

\bibitem[{{Mazzali} {et~al.}(1998){Mazzali}, {Cappellaro}, {Danziger},
  {Turatto}, \& {Benetti}}]{Mazzali98}
{Mazzali}, P.~A., {Cappellaro}, E., {Danziger}, I.~J., {Turatto}, M., \&
  {Benetti}, S. 1998, \apjl, 499, L49

\bibitem[{{Milisavljevic} {et~al.}(2014){Milisavljevic}, {Margutti},
  {Crabtree}, {Foster}, {Soderberg}, {Fesen}, {Parrent}, {Sanders}, {Drout},
  {Kamble}, {Chakraborti}, {Pickering}, {Cenko}, {Silverman}, {Filippenko},
  {Kirshner}, {Mazzali}, {Maeda}, {Marion}, {Vinko}, \&
  {Wheeler}}]{Milisavljevic14}
{Milisavljevic}, D., {Margutti}, R., {Crabtree}, K.~N., {et~al.} 2014, \apjl,
  782, L5

\bibitem[{{Modigliani} {et~al.}(2010){Modigliani}, {Goldoni}, {Royer},
  {Haigron}, {Guglielmi}, {Fran{\c c}ois}, {Horrobin}, {Bristow}, {Vernet},
  {Moehler}, {Kerber}, {Ballester}, {Mason}, \& {Christensen}}]{Modigliani10}
{Modigliani}, A., {Goldoni}, P., {Royer}, F., {et~al.} 2010, in Society of
  Photo-Optical Instrumentation Engineers (SPIE) Conference Series, Vol. 7737,
  Society of Photo-Optical Instrumentation Engineers (SPIE) Conference Series

\bibitem[{{Nomoto}(1982)}]{Nomoto82}
{Nomoto}, K. 1982, \apj, 253, 798

\bibitem[{{Nugent} {et~al.}(1995){Nugent}, {Phillips}, {Baron}, {Branch}, \&
  {Hauschildt}}]{Nugent95}
{Nugent}, P., {Phillips}, M., {Baron}, E., {Branch}, D., \& {Hauschildt}, P.
  1995, \apjl, 455, L147

\bibitem[{{O'Donnell}(1994)}]{ODonnell94}
{O'Donnell}, J.~E. 1994, \apj, 437, 262

\bibitem[{{Pakmor} {et~al.}(2012){Pakmor}, {Kromer}, {Taubenberger}, {Sim},
  {R{\"o}pke}, \& {Hillebrandt}}]{Pakmor12}
{Pakmor}, R., {Kromer}, M., {Taubenberger}, S., {et~al.} 2012, \apjl, 747, L10

\bibitem[{{Park} {et~al.}(2015){Park}, {Jeong}, \& {Yi}}]{Park15}
{Park}, J., {Jeong}, H., \& {Yi}, S.~K. 2015, ArXiv e-prints

\bibitem[{{Pastorello} {et~al.}(2007)}]{Pastorello07}
{Pastorello}, A. {et~al.} 2007, \mnras, 376, 1301

\bibitem[{{Patat} {et~al.}(2009){Patat}, {Baade}, {H{\"o}flich}, {Maund},
  {Wang}, \& {Wheeler}}]{Patat09}
{Patat}, F., {Baade}, D., {H{\"o}flich}, P., {et~al.} 2009, \aap, 508, 229

\bibitem[{{Patat} {et~al.}(1996){Patat}, {Benetti}, {Cappellaro}, {Danziger},
  {della Valle}, {Mazzali}, \& {Turatto}}]{Patat96}
{Patat}, F., {Benetti}, S., {Cappellaro}, E., {et~al.} 1996, \mnras, 278, 111

\bibitem[{{Patat} {et~al.}(2007){Patat}, {Chandra}, {Chevalier}, {Justham},
  {Podsiadlowski}, {Wolf}, {Gal-Yam}, {Pasquini}, {Crawford}, {Mazzali},
  {Pauldrach}, {Nomoto}, {Benetti}, {Cappellaro}, {Elias-Rosa}, {Hillebrandt},
  {Leonard}, {Pastorello}, {Renzini}, {Sabbadin}, {Simon}, \&
  {Turatto}}]{Patat07}
{Patat}, F., {Chandra}, P., {Chevalier}, R., {et~al.} 2007, Science, 317, 924

\bibitem[{{Phillips}(1993)}]{Phillips93}
{Phillips}, M.~M. 1993, \apjl, 413, L105

\bibitem[{{Phillips} {et~al.}(1999){Phillips}, {Lira}, {Suntzeff}, {Schommer},
  {Hamuy}, \& {Maza}}]{Phillips99}
{Phillips}, M.~M., {Lira}, P., {Suntzeff}, N.~B., {et~al.} 1999, \aj, 118, 1766

\bibitem[{{Phillips} {et~al.}(2013)}]{Phillips13}
{Phillips}, M.~M. {et~al.} 2013, \apj, 779, 38

\bibitem[{{Pignata} {et~al.}(2008){Pignata}, {Benetti}, {Mazzali}, {Kotak},
  {Patat}, {Meikle}, {Stehle}, {Leibundgut}, {Suntzeff}, {Buson}, {Cappellaro},
  {Clocchiatti}, {Hamuy}, {Maza}, {Mendez}, {Ruiz-Lapuente}, {Salvo},
  {Schmidt}, {Turatto}, \& {Hillebrandt}}]{Pignata08}
{Pignata}, G., {Benetti}, S., {Mazzali}, P.~A., {et~al.} 2008, \mnras, 388, 971

\bibitem[{{Pignata} {et~al.}(2010){Pignata}, {Cifuentes}, {Maza}, {Hamuy},
  {Antezana}, {Gonzalez}, {Gonzalez}, {Silva}, {Folatelli}, {Cartier},
  {Forster}, {Marchi}, {Conuel}, {Reichart}, {Ivarsen}, {Haislip}, {Crain},
  {Foster}, {Nysewander}, \& {Lacluyze}}]{Pignata10}
{Pignata}, G., {Cifuentes}, M., {Maza}, J., {et~al.} 2010, Central Bureau
  Electronic Telegrams, 2344, 1

\bibitem[{{Pignata} {et~al.}(2004)}]{Pignata04}
{Pignata}, G. {et~al.} 2004, \mnras, 355, 178

\bibitem[{{Poznanski} {et~al.}(2011){Poznanski}, {Ganeshalingam}, {Silverman},
  \& {Filippenko}}]{Poznanski11}
{Poznanski}, D., {Ganeshalingam}, M., {Silverman}, J.~M., \& {Filippenko},
  A.~V. 2011, \mnras, 415, L81

\bibitem[{{Poznanski} {et~al.}(2012){Poznanski}, {Prochaska}, \&
  {Bloom}}]{Poznanski12}
{Poznanski}, D., {Prochaska}, J.~X., \& {Bloom}, J.~S. 2012, \mnras, 426, 1465

\bibitem[{{Raskin} {et~al.}(2013){Raskin}, {Kasen}, {Moll}, {Schwab}, \&
  {Woosley}}]{Raskin13}
{Raskin}, C., {Kasen}, D., {Moll}, R., {Schwab}, J., \& {Woosley}, S. 2013,
  ArXiv e-prints

\bibitem[{{Raskin} {et~al.}(2009){Raskin}, {Timmes}, {Scannapieco}, {Diehl}, \&
  {Fryer}}]{Raskin09}
{Raskin}, C., {Timmes}, F.~X., {Scannapieco}, E., {Diehl}, S., \& {Fryer}, C.
  2009, \mnras, 399, L156

\bibitem[{{Riess} {et~al.}(1996){Riess}, {Press}, \& {Kirshner}}]{Riess96}
{Riess}, A.~G., {Press}, W.~H., \& {Kirshner}, R.~P. 1996, \apj, 473, 588

\bibitem[{{Scalzo} {et~al.}(2014{\natexlab{a}}){Scalzo}, {Aldering},
  {Antilogus}, {Aragon}, {Bailey}, {Baltay}, {Bongard}, {Buton},
  {Cellier-Holzem}, {Childress}, {Chotard}, {Copin}, {Fakhouri}, {Gangler},
  {Guy}, {Kim}, {Kowalski}, {Kromer}, {Nordin}, {Nugent}, {Paech}, {Pain},
  {Pecontal}, {Pereira}, {Perlmutter}, {Rabinowitz}, {Rigault}, {Runge},
  {Saunders}, {Sim}, {Smadja}, {Tao}, {Taubenberger}, {Thomas}, {Weaver}, \&
  {Nearby Supernova Factory}}]{Scalzo14a}
{Scalzo}, R., {Aldering}, G., {Antilogus}, P., {et~al.} 2014{\natexlab{a}},
  \mnras, 440, 1498

\bibitem[{{Scalzo} {et~al.}(2014{\natexlab{b}}){Scalzo}, {Ruiter}, \&
  {Sim}}]{Scalzo14b}
{Scalzo}, R.~A., {Ruiter}, A.~J., \& {Sim}, S.~A. 2014{\natexlab{b}}, \mnras,
  445, 2535

\bibitem[{{Schlafly} \& {Finkbeiner}(2011)}]{Schlafly11}
{Schlafly}, E.~F. \& {Finkbeiner}, D.~P. 2011, \apj, 737, 103

\bibitem[{{Schlegel} {et~al.}(1998){Schlegel}, {Finkbeiner}, \&
  {Davis}}]{Schlegel98}
{Schlegel}, D.~J., {Finkbeiner}, D.~P., \& {Davis}, M. 1998, \apj, 500, 525

\bibitem[{{Shen} {et~al.}(2013){Shen}, {Guillochon}, \& {Foley}}]{Shen13}
{Shen}, K.~J., {Guillochon}, J., \& {Foley}, R.~J. 2013, \apjl, 770, L35

\bibitem[{{Silverman} \& {Filippenko}(2012)}]{Silverman12c}
{Silverman}, J.~M. \& {Filippenko}, A.~V. 2012, \mnras, 425, 1917

\bibitem[{{Silverman} {et~al.}(2013){Silverman}, {Ganeshalingam}, \&
  {Filippenko}}]{Silverman13}
{Silverman}, J.~M., {Ganeshalingam}, M., \& {Filippenko}, A.~V. 2013, \mnras,
  430, 1030

\bibitem[{{Silverman} {et~al.}(2012){Silverman}, {Kong}, \&
  {Filippenko}}]{Silverman12a}
{Silverman}, J.~M., {Kong}, J.~J., \& {Filippenko}, A.~V. 2012, \mnras, 425,
  1819

\bibitem[{{Sim} {et~al.}(2012){Sim}, {Fink}, {Kromer}, {R{\"o}pke}, {Ruiter},
  \& {Hillebrandt}}]{Sim12}
{Sim}, S.~A., {Fink}, M., {Kromer}, M., {et~al.} 2012, \mnras, 420, 3003

\bibitem[{{Simon} {et~al.}(2009){Simon}, {Gal-Yam}, {Gnat}, {Quimby},
  {Ganeshalingam}, {Silverman}, {Blondin}, {Li}, {Filippenko}, {Wheeler},
  {Kirshner}, {Patat}, {Nugent}, {Foley}, {Vogt}, {Butler}, {Peek},
  {Rosolowsky}, {Herczeg}, {Sauer}, \& {Mazzali}}]{Simon09}
{Simon}, J.~D., {Gal-Yam}, A., {Gnat}, O., {et~al.} 2009, \apj, 702, 1157

\bibitem[{{Smith} {et~al.}(2002){Smith}, {Tucker}, {Kent}, {Richmond},
  {Fukugita}, {Ichikawa}, {Ichikawa}, {Jorgensen}, {Uomoto}, {Gunn}, {Hamabe},
  {Watanabe}, {Tolea}, {Henden}, {Annis}, {Pier}, {McKay}, {Brinkmann}, {Chen},
  {Holtzman}, {Shimasaku}, \& {York}}]{Smith02}
{Smith}, J.~A., {Tucker}, D.~L., {Kent}, S., {et~al.} 2002, \aj, 123, 2121

\bibitem[{{Stanishev} {et~al.}(2007){Stanishev}, {Goobar}, {Benetti}, {Kotak},
  {Pignata}, {Navasardyan}, {Mazzali}, {Amanullah}, {Garavini}, {Nobili},
  {Qiu}, {Elias-Rosa}, {Ruiz-Lapuente}, {Mendez}, {Meikle}, {Patat},
  {Pastorello}, {Altavilla}, {Gustafsson}, {Harutyunyan}, {Iijima},
  {Jakobsson}, {Kichizhieva}, {Lundqvist}, {Mattila}, {Melinder}, {Pavlenko},
  {Pavlyuk}, {Sollerman}, {Tsvetkov}, {Turatto}, \&
  {Hillebrandt}}]{Stanishev07}
{Stanishev}, V., {Goobar}, A., {Benetti}, S., {et~al.} 2007, \aap, 469, 645

\bibitem[{{Stehle} {et~al.}(2005){Stehle}, {Mazzali}, {Benetti}, \&
  {Hillebrandt}}]{Stehle05}
{Stehle}, M., {Mazzali}, P.~A., {Benetti}, S., \& {Hillebrandt}, W. 2005,
  \mnras, 360, 1231

\bibitem[{{Sternberg} {et~al.}(2011){Sternberg}, {Gal-Yam}, {Simon}, {Leonard},
  {Quimby}, {Phillips}, {Morrell}, {Thompson}, {Ivans}, {Marshall},
  {Filippenko}, {Marcy}, {Bloom}, {Patat}, {Foley}, {Yong}, {Penprase},
  {Beeler}, {Allende Prieto}, \& {Stringfellow}}]{Sternberg11}
{Sternberg}, A., {Gal-Yam}, A., {Simon}, J.~D., {et~al.} 2011, Science, 333,
  856

\bibitem[{{Sternberg} {et~al.}(2013){Sternberg}, {Gal Yam}, {Simon}, {Patat},
  {Hillebrandt}, {Phillips}, {Foley}, {Thompson}, {Morrell}, {Chomiuk},
  {Soderberg}, {Yong}, {Kraus}, {Herczeg}, {Hsiao}, {Raskutti}, {Cohen},
  {Mazzali}, \& {Nomoto}}]{Sternberg13}
{Sternberg}, A., {Gal Yam}, A., {Simon}, J.~D., {et~al.} 2013, ArXiv e-prints

\bibitem[{{Stritzinger}(2010)}]{Stritzinger10ev}
{Stritzinger}, M. 2010, Central Bureau Electronic Telegrams, 2346, 1

\bibitem[{{Stritzinger} {et~al.}(2006){Stritzinger}, {Leibundgut}, {Walch}, \&
  {Contardo}}]{Stritzinger06a}
{Stritzinger}, M., {Leibundgut}, B., {Walch}, S., \& {Contardo}, G. 2006, \aap,
  450, 241

\bibitem[{{Tripp}(1998)}]{Tripp98}
{Tripp}, R. 1998, \aap, 331, 815

\bibitem[{{Turatto} {et~al.}(2003){Turatto}, {Benetti}, \&
  {Cappellaro}}]{Turatto03}
{Turatto}, M., {Benetti}, S., \& {Cappellaro}, E. 2003, in From Twilight to
  Highlight: The Physics of Supernovae, ed. W.~{Hillebrandt} \&
  B.~{Leibundgut}, 200

\bibitem[{{Valenti} {et~al.}(2008)}]{Valenti08}
{Valenti}, S. {et~al.} 2008, \mnras, 383, 1485

\bibitem[{{Wang}(2005)}]{Wang05}
{Wang}, L. 2005, \apjl, 635, L33

\bibitem[{{Wang} {et~al.}(2003)}]{Wang03a}
{Wang}, L. {et~al.} 2003, \apj, 591, 1110

\bibitem[{{Wang} {et~al.}(2009){Wang}, {Filippenko}, {Ganeshalingam}, {Li},
  {Silverman}, {Wang}, {Chornock}, {Foley}, {Gates}, {Macomber}, {Serduke},
  {Steele}, \& {Wong}}]{Wang09}
{Wang}, X., {Filippenko}, A.~V., {Ganeshalingam}, M., {et~al.} 2009, \apjl,
  699, L139

\bibitem[{{Wang} {et~al.}(2008)}]{Wang08}
{Wang}, X. {et~al.} 2008, \aj, 626-643, 1

\bibitem[{{Webbink}(1984)}]{Webbink84}
{Webbink}, R.~F. 1984, \apj, 277, 355

\bibitem[{{Welty} {et~al.}(2014){Welty}, {Ritchey}, {Dahlstrom}, \&
  {York}}]{Welty14}
{Welty}, D.~E., {Ritchey}, A.~M., {Dahlstrom}, J.~A., \& {York}, D.~G. 2014,
  ArXiv e-prints

\bibitem[{{Wheeler} {et~al.}(1998){Wheeler}, {Hoeflich}, {Harkness}, \&
  {Spyromilio}}]{Wheeler98}
{Wheeler}, J.~C., {Hoeflich}, P., {Harkness}, R.~P., \& {Spyromilio}, J. 1998,
  \apj, 496, 908

\bibitem[{{Zelaya} {et~al.}(2015)}]{Zelaya15}
{Zelaya}, P. {et~al.} 2015, in preparation

\end{thebibliography}
\end{document}